\documentclass{article} 

\usepackage{caption}
\usepackage{subcaption}
\usepackage{jmlr2e}
\usepackage{hyperref}
\usepackage{url}
\usepackage{drago}
\usepackage[on]{editing}
\usepackage{courier}

\title{Interaction Testing in Variation Analysis}

\author{Drago Ple\v cko \\
    \addr{
        Department of Computer Science\\
        Columbia University\\
        New York, NY 10027, USA \\
        \texttt{dp3144@columbia.edu}
    }
}

\usepackage{selectp}

\begin{document}

\maketitle


\begin{abstract}
Relationships of cause and effect are of prime importance for explaining scientific phenomena. Often, rather than just understanding the effects of causes, researchers also wish to understand how a cause $X$ affects an outcome $Y$ mechanistically -- i.e., what are the causal pathways that are activated between $X$ and $Y$. For analyzing such questions, a range of methods has been developed over decades under the rubric of causal mediation analysis. Traditional mediation analysis focuses on decomposing the average treatment effect (ATE) into direct and indirect effects, and therefore focuses on the ATE as the central quantity. This corresponds to providing explanations for associations in the interventional regime, such as when the treatment $X$ is randomized. 
Commonly, however, it is of interest to explain associations in the natural, observational regime, and not just in the interventional regime. In this paper, we introduce \text{variation analysis}, an extension of mediation analysis that focuses on the total variation (TV) measure between $X$ and $Y$, written as $\ex[Y \mid X=x_1] - \ex[Y \mid X=x_0]$.  The TV measure encompasses both causal and confounded effects, as opposed to the ATE which only encompasses causal (direct and mediated) variations. In this way, the TV measure is suitable for providing explanations in the natural regime and answering questions such as ``why is $X$ associated with $Y$ in a particular way?''. 
Our focus is on decomposing the TV measure, in a way that explicitly includes direct, indirect, and confounded variations. Furthermore, we also decompose the TV measure to include interaction terms between these different pathways. 
Subsequently, the concept of interaction testing is introduced, which involves hypothesis tests to determine if interaction terms are significantly different from zero. If interactions are not significant, a more parsimonious decomposition of the TV measure can be used. 
The paper further provides a structural basis for these interaction tests (through the language of structural causal models) and demonstrates their applicability through synthetic and real-world data analyses. The extension of the framework for log-risk and log-odds scales for binary outcomes is also discussed, offering a comprehensive approach to understanding the interplay of direct, indirect, and confounded effects in causal inference.
\end{abstract}

\section{Introduction}
Understanding relationships of cause and effect is one of the fundamental tasks found throughout the sciences. The process of establishing mechanistic links between causes and their consequences is at the core of our ability to explain why events occur as they do. In this context, mediation analysis, a widely used  tool, helps unravel the pathways through which causal effects are transmitted. By identifying intermediary variables, mediation analysis offers deeper insights into the underlying mechanisms driving the observed cause-effect relationships. This approach is crucial in fields ranging from epidemiology to social sciences, where understanding the nuances of causal relationships can inform interventions and policy decisions.

Interestingly, most of the literature on mediation analysis focuses on understanding the variations contained in the average treatment effect (ATE), also known as the total effect (TE), given by
\begin{align}
    \ex[Y \mid do(X = x_1)] - \ex[Y \mid do(X = x_0)], 
\end{align}
where $do(\cdot)$ symbolizes the do-operator \citep{pearl:2k}, and $x_0, x_1$ are two distinct values attained by the binary variable $X$.
Instead of just quantifying the causal effect through the ATE and the related quantities, researchers are often more broadly interested in determining which causal mechanisms transmit the causal influences from $X$ to $Y$. Various approaches for solving this problem have been proposed under the rubric of causal mediation analysis, and the associated literature is vast \citep{baron1986moderator, robins1992identifiability, pearl:01, imai2010general, vanderweele2015explanation}. A common goal for many of the mediation methods is to \textit{decompose variations} that are contained in the ATE into variations that are mediated by other variables (known as the indirect or mediated effect) and variations that are not mediated by other variables (known as the direct effect). 

Interestingly, mediation analysis focuses solely on causal variations, and the ATE captures the association of $X$ and $Y$ in an interventional regime, such as the randomized control trial (RCT), in which values of $X$ are randomized to either $x_0$ or $x_1$. This approach has proven tremendously useful for explaining causal effects, such as in testing the effects of drugs, understanding the impact of educational interventions, and evaluating policy changes. 
Often, however, researchers may be interested in explaining the association of $X$ and $Y$ in the natural, observational regime, without a specific intervention in mind. Somewhat surprisingly, the common approach for mediation falls short of answering simple questions such as ``why do patients receiving chemotherapy have higher mortality rates than those not receiving it?'', or ``why do coffee drinkers have higher rates of cardiovascular disease?''.
In both of the examples, the causal relationship may account for only a part of the observed association, while non-causal (or spurious/confounded) effects also play an important role in explaining the phenomenon. In the former example, illness severity increases both the probability of receiving chemotherapy and dying, while in the latter, coffee drinkers are more likely to also smoke, which is a known determinant for cardiovascular disease.
When explaining associations in the natural, observational regime, we may be interested in the quantity
\begin{align}
    \ex[Y \mid X = x_1] - \ex[Y \mid X = x_0],
\end{align}
which we will refer to as the total variation (TV) measure, instead of the typically used ATE. The key difference between the ATE and the TV is that the latter encompasses confounded variations and not just the causal ones. As the approaches focusing on the ATE were called mediation analysis, in this paper, we call the approaches focusing on the TV measure \textit{variation analysis}, to provide a distinction for a class of methods also concerned with effects that are neither direct nor mediated, but confounded.

While mediation analysis has been successful in quantifying direct and indirect effects, far fewer approaches for handling \textit{interactions} between direct and indirect effects have been proposed \citep{vanderweele2015explanation}. Furthermore, in the context of variation analysis, no methods exist to date that are intended for analyzing interactions, such as the interactions of spurious and causal effects. In the following example, we motivate some of the key developments appearing in this paper: 
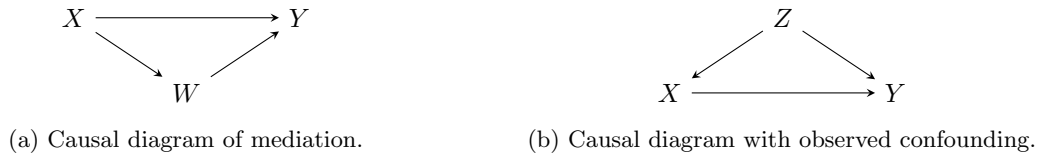
\begin{figure}
     \centering
     \begin{subfigure}[b]{0.48\textwidth}
         \centering
     \scalebox{1}{
     \begin{tikzpicture}
	 [>=stealth, rv/.style={thick}, rvc/.style={triangle, draw, thick, minimum size=7mm}, node distance=18mm]
	 \pgfsetarrows{latex-latex};
	 	\node[rv] (x) at (-1.5,0) {$X$};
	 	\node[rv] (w) at (0,-1) {$W$};
            \node[rv] (y) at (1.5,0) {$Y$};
	 	
	 	\draw[->] (x) -- (y);
            \draw[->] (x) -- (w);
            \path[->] (w) edge[bend left = -0] (y);
	 \end{tikzpicture}
     }   
     \caption{Causal diagram of mediation.}
     \label{fig:med-graph}
     \end{subfigure}
     \hfill
     \begin{subfigure}[b]{0.48\textwidth}
     \centering
    \scalebox{1}{
     \begin{tikzpicture}
	 [>=stealth, rv/.style={thick}, rvc/.style={triangle, draw, thick, minimum size=7mm}, node distance=18mm]
	 \pgfsetarrows{latex-latex};
	 	\node[rv] (x) at (-1.5,0) {$X$};
	 	\node[rv] (z) at (0,1) {$Z$};
            \node[rv] (y) at (1.5,0) {$Y$};
	 	
	 	\draw[->] (x) -- (y);
            \draw[->] (z) -- (y);
            \path[->] (z) edge[bend left = -0] (x);
	 \end{tikzpicture}
     } 
    \caption{Causal diagram with observed confounding.}
    \label{fig:conf-graph}
     \end{subfigure}
     \caption{Causal diagrams for Ex.~\ref{ex:intro-decompositions}.}
\end{figure}
\begin{example}[Total Effect and Total Variation Decompositions] \label{ex:intro-decompositions}
    Consider the causal diagram in Fig.~\ref{fig:med-graph}, with treatment $X$, outcome $Y$, and mediator $W$. A key result from \cite{pearl:01} demonstrated that the average treatment effect (ATE) can be decomposed into
    \begin{align} \label{eq:pearl-mediation-formula}
         \ex[Y \mid do(X = x_1)] - \ex[Y \mid do(X = x_0)] = \underbrace{\ex[Y_{x_1, W_{x_0}} - Y_{x_0, W_{x_0}}]}_{\text{natural direct effect}} - \underbrace{\ex[Y_{x_1, W_{x_0}} - Y_{x_1, W_{x_1}}]}_{\text{natural indirect effect}}.
    \end{align}
    The natural direct effect (NDE) compares the potential outcome $Y_{x_0, W_{x_0}}$ in which $X = x_0$ along both direct ($X \to Y$) and indirect ($X \to W \to Y$) pathways, vs. the potential outcome $Y_{x_1, W_{x_0}}$ in which $X = x_1$ along the direct pathway, while $W$ varies naturally at the level of $X = x_0$, written $W_{x_0}$. In this way, the NDE quantifies the effect of changing $X$ from $x_0$ to $x_1$ along the direct pathway. To obtain the ATE, from the NDE, one subtracts the natural indirect effect (NIE) with a reverse transition, which measures the effect of changing $X = x_1$ to $X = x_0$ along the indirect path by comparing potential outcomes $Y_{x_1, W_{x_1}}$ vs. $Y_{x_1, W_{x_0}}$ which respond to $X = x_1$ along the direct path. Often, the need to consider a transition $x_0 \to x_1$ in the NDE while subtracting a reverse transition $x_1 \to x_0$ for the NIE is criticized as a shortcoming of the widely used effect decomposition in Eq.~\ref{eq:pearl-mediation-formula}.

    A result of \cite{vanderweele2015explanation} sheds lights on the issue of opposite transitions appearing in the decomposition of the ATE. In particular, the ATE can also be decomposed in a different way:
    \begin{align} \label{eq:tyler-decomp}
        \text{ATE}_{x_0, x_1}(y) &= \underbrace{\ex[Y_{x_1, W_{x_0}} - Y_{x_0, W_{x_0}}]}_{\text{natural direct effect}} + \underbrace{\ex[Y_{x_0, W_{x_1}} - Y_{x_0, W_{x_0}}]}_{\text{natural indirect effect}} \\
        &+ \underbrace{\ex[Y_{x_1, W_{x_0}} - Y_{x_0, W_{x_0}} - (Y_{x_1, W_{x_1}} - Y_{x_0, W_{x_1}})]}_{\text{interaction effect}}.
    \end{align}
    Notably, in Eq.~\ref{eq:tyler-decomp}, the NDE and the NIE both appear with a transition of $x_0 \to x_1$, with a baseline of $Y_{x_0, W_{x_0}}$. There is an additional term, however, which compares how the direct effect changes according to the behavior of $W$, namely
    \begin{align}
        \ex[\underbrace{Y_{x_1, W_{x_0}} - Y_{x_0, W_{x_0}}}_{x_0\to x_1\text{ DE with } W_{x_0}} - \underbrace{(Y_{x_1, W_{x_1}} - Y_{x_0, W_{x_1}})}_{x_0\to x_1\text{ DE with } W_{x_1}}].
    \end{align}
    In other words, the interactive effect compares how strongly the direct effect of a $x_0 \to x_1$ transition changes in the setting of $W_{x_0}$ vs. $W_{x_1}$. The result of \cite{vanderweele2015explanation} demonstrates that the key issue in interpreting Pearl's decomposition (the appearance of effects with opposite transitions in $X$) is fundamentally due to interactions.

    The total variation (TV) measure, when considering the causal diagram in Fig.~\ref{fig:conf-graph}, can be decomposed as \citep{zhang2018fairness, plevcko2024causal}:
    \begin{align}
        \hspace{-2pt}\ex[Y \mid X = x_1] - \ex[Y \mid X = x_0] = \underbrace{\ex[Y_{x_1} - Y_{x_0} \mid X = x_0]}_{x_0\text{-specific total effect}} - \underbrace{\ex[Y_{x_1} \mid X = x_0] - \ex[Y_{x_1} \mid X=x_1]}_{x\text{-specific spurious effect}}.
    \end{align}
    The $x_0$-specific total effect (also known as the effect of treatment on the treated, ETT) computes the effect of a $x_0 \to x_1$ transition for the subpopulation of individuals with $X = x_0$. To obtain the TV measure, from this effect we subtract the $x$-specific spurious effect which compares how conditioning on $X = x_0$ differs from $X = x_1$ while setting $X = x_1$ along the causal pathways. Similarly to Eq.~\ref{eq:pearl-mediation-formula}, the two effects appear with reverse transitions.

    The problem can be remedied again, by noting that the TV measure can also be decomposed as:
    \begin{align}
        \text{TV}_{x_0, x_1}(y) &= \underbrace{\ex[Y_{x_1} - Y_{x_0} \mid X = x_0]}_{x_0\text{-specific total effect}} + \underbrace{\ex[Y_{x_0} \mid X = x_1] - \ex[Y_{x_0} \mid X = x_0]}_{x\text{-specific spurious effect}} \\
        &+ \underbrace{\ex[Y_{x_1} - Y_{x_0} \mid X = x_1] - \ex[Y_{x_1} - Y_{x_0} \mid X = x_0]}_{\text{causal/spurious interaction}}.
    \end{align}
    The additional term appearing in this new decomposition compares
    \begin{align}
        \underbrace{\ex[Y_{x_1} - Y_{x_0} \mid X = x_1]}_{x_0 \to x_1\text{ TE with } X = x_1 \text{ conditioning}} \text{ vs. } \underbrace{\ex[Y_{x_1} - Y_{x_0} \mid X = x_0]}_{x_0 \to x_1 \text{ TE with } X = x_0 \text{ conditioning}}, \label{eq:intro-TE-SE}
    \end{align}
    and quantifies how much the total effect of a $x_0 \to x_1$ transition changes when conditioning on $X = x_1$ vs. $X = x_0$. In this way, we can measure the strength of the interaction between spurious and causal paths. 
\end{example}
Our goal in this manuscript will be to provide a coherent umbrella for understanding interactions of causal pathways in variation analysis, namely direct, indirect, and spurious. The contributions of the paper can be summarized as follows:
\begin{enumerate}[label=(\roman*)]
    \item We prove a first decomposition of the total variation (TV) measure that contains an explicit term for the interaction of causal and spurious pathways (Thm.~\ref{thm:1st-tv-decomp}),
    \item We develop the concept of \textit{interaction testing} (Def.~\ref{def:ia-test}), in which an explicit interaction term that appears in a decomposition is subject to a hypothesis test of being equal to $0$,
    \item We develop an algorithm for non-parametric interaction testing  (Alg.~\ref{algo:ia-testing}), showing a common preference for parsimony in statistics: if the interaction term is not significantly different from $0$, a more parsimonious TV decomposition may be used. 
    \item We relate the effect interactions to the structural causal mechanisms of the underlying system (Def.~\ref{def:str-ia}), and demonstrate when there is a correspondence between a mechanism property and the corresponding interaction test (Prop.~\ref{prop:str-admissibility}),
    \item We provide the most general decomposition of the TV measure that provides accounts for all interactions between direct, indirect, and spurious effects (Thm.~\ref{thm:2nd-tv-decomp}),
    \item We translate our results to the log-risk and log-odds scales (Appendix~\ref{appendix:lr-lo-scales}),
    \item We perform an in-depth empirical analysis with the attempt to discover how commonly effects interact in practice (Sec.~\ref{sec:experiments}).
\end{enumerate}
\section{Preliminaries}
We use the language of structural causal models (SCMs) as our basic semantical framework \citep{pearl:2k}. A structural causal model (SCM) is defined as:
\begin{definition}[Structural Causal Model (SCM) \citep{pearl:2k}] \label{def:SCM}
	A structural causal model $\mathcal{M}$ is a 4-tuple $\langle V, U, \mathcal{F}, \pr(U)\rangle$, where
  \begin{enumerate}
    \item $U$ is a set of exogenous variables, also called background variables, that are determined by factors outside the model;
    \item $V = \lbrace V_1, ..., V_n \rbrace$ is a set of endogenous (observed) variables, that are determined by variables in the model (i.e. by the variables in $U \cup V$);
    \item $\mathcal{F} = \lbrace f_{V_1}, ..., f_{V_n} \rbrace$ is the set of structural functions determining $V$, $v_i \gets f_{V_i}(\pa(v_i), u_i)$, where $\pa(V_i) \subseteq V \setminus V_i$ and $U_i \subseteq U$ are the functional arguments of $f_{V_i}$;
    \item $\pr(U)$ is a distribution over the exogenous variables $U$.
  \end{enumerate}
\end{definition}
\noindent The assignment mechanisms $\mathcal{F}$ determine how each of the observed variables $V_i$ attains its value, based on other observed variables and the latent variables $U$. Together with the probability distribution $\pr(U)$ over the exogenous variables $U$, it specifies the entire behavior of the underlying phenomenon. In many places, expectations are written as simple summations, whereas for continuous variables, such notation should be replaced with integrals, assuming that the corresponding probability measure admits a density w.r.t. the Lebesgue measure. For notational simplicity, we often use shorthand notation, with expressions such as $\pr(Y = y)$ replaced with $\pr (y)$, whenever the simplification is clear from the context.

Crucially, the SCM induces the \textit{observational distribution} of the underlying phenomenon, defined through:
\begin{definition}[Observational Distribution \citep{bareinboim2020on}] \label{def:obs-dist}
An SCM $\mathcal{M}$ that is a 4-tuple $\langle V, U, \mathcal{F}, \pr(U) \rangle$ induces a joint probability distribution $\pr(V)$ such that for each $Y \subseteq V$,
\begin{align}
    \pr^{\mathcal{M}}(y) = \sum_{u} \mathbb{1}\Big(Y(u) = y \Big) \pr(u),
\end{align} 
where $Y(u)$ is the solution for $Y$ after evaluating $\mathcal{F}$ with $U = u$.
\end{definition}
\noindent A further important notion building on the concept of the SCM is that of a submodel, which is defined next:
\begin{definition}[Submodel \citep{pearl:2k}] \label{def:submodel}
    Let $\mathcal{M}$ be a structural causal model, $X$ a set of variables in $V$, and $x$ a particular value of $X$. A submodel $\mathcal{M}_{x}$ (of $\mathcal{M}$) is a 4-tuple:
    \begin{equation}
        \mathcal{M}_{x} = \langle V, U, \mathcal{F}_{x}, \pr(U)\rangle
    \end{equation}
    where 
    \begin{equation}
        \mathcal{F}_{x} = \lbrace f_i : V_i \notin X \rbrace \cup \lbrace X \gets x\rbrace,
    \end{equation}
    and all other components are preserved from $\mathcal{M}$. 
\end{definition}
\noindent Building on submodels, we introduce next the
notion of a potential outcome:
\begin{definition}[Potential Outcome / Response \citep{rubin1974estimating, pearl:2k}]\label{def:potentialresponse}
    Let $X$ and $Y$ be two sets of variables in $V$ and $u \in \mathcal{U}$ be a unit. The potential outcome/response $Y_x(u)$ is defined as the solution for $Y$ of the set of equations $\mathcal{F}_x$ evaluated with $U=u$. That is, $Y_x(u)$ denotes the solution of $Y$ in the submodel $\mathcal{M}_x$ of $\mathcal{M}$.
\end{definition}
\noindent In words, $Y_x(u)$ is the value variable $Y$ would take if (possibly contrary to observed facts) $X$ is set to $x$, for a specific unit $u$. We further define how counterfactual distributions over various possible potential outcomes are computed:
\begin{definition}[Counterfactual Distributions \citep{bareinboim2020on}] \label{def:ctf-dist}
    Consider an SCM $\mathcal{M} = \langle V, U, \mathcal{F}, \pr(u) \rangle$, and let $Y_1, \dots, Y_k \subset V$, and $X_1, \dots, X_k \subset V$ be subsets of the observables, and let $x_1, \dots, x_k$ be specific values of $X_i$s. Denote by $(Y_i)_{x_i}$ the potential response of variables $Y_i$ when setting $X_i = x_i$. The SCM $\mathcal{M}$ induces a family of joint distributions over counterfactual events $(Y_1)_{x_1}, \dots, (Y_k)_{x_k}$:
    \begin{align} \label{eq:L3def}
        \pr^{\mathcal{M}}((y_1)_{x_1}, \dots, (y_k)_{x_k}) = \sum_{u} \mathbb{1}\Big( \bigwedge_{i=1}^k (Y_i)_{x_i}(u) = y_i \Big) \pr(u).
    \end{align}
\end{definition}
\noindent The LHS in Eq.~\ref{eq:L3def} contains variables with different subscripts, which syntactically represent different potential responses (Def.~\ref{def:potentialresponse}), or counterfactual worlds. 
Finally, there is one more prerequisite notion for our discussion. The mechanisms $\mathcal{F}$ and the distribution over the exogenous variables $\pr(U)$ are almost never observed. However, to perform causal inference, we need a way of encoding assumptions about the underlying SCM. A common way of doing so is through an object called a causal diagram, which is defined next: 
\begin{definition}[Causal Diagram \citep{pearl:2k, bareinboim2020on}] \label{def:diagram}
	Let an SCM $\mathcal{M}$ be a 4-tuple $\langle V, U, \mathcal{F}, \pr(U)\rangle$. A graph $\mathcal{G}$ is said to be a \textit{causal diagram} (of $\mathcal{M}$) if:
	\begin{enumerate}[label=(\arabic*)]
		\item there is a vertex for every endogenous variable $V_i \in V$,
		\item there is an edge $V_i \to V_j$ if $V_i$ appears as an argument of $f_j \in \mathcal{F}$,
		\item there is a bidirected edge $V_i\dashleftarrow\dasharrow V_j$ if the corresponding $U_i, U_j \subset U$ are correlated or the corresponding functions $f_i, f_j$ share some $U_{ij} \in U$ as an argument.
	\end{enumerate} \vspace{-0.3in}
\end{definition}
We call $\pa(V_i)$ the set of parents of $V_i$,  while the sets of children $\ch(V_i)$, ancestors $\an(V_i)$, and descendants $\de(V_i)$ are defined analogously.

Building on the notion of a potential response, one can further define the notions of counterfactual and factual contrasts, given by:
\begin{definition}[Contrasts \citep{plevcko2024causal}] \label{def:contrast}
Given an SCM $\mathcal{M}$,  a contrast $\mathcal{C}$ is any quantity of the form 
\begin{equation} \label{eq:contrast}
    \mathcal{C}(C_0, C_1, E_0, E_1) = \ex[Y_{C_1} \mid E_1] - \ex[Y_{C_0}\mid E_0],
\end{equation}
where $E_0, E_1$ are observed (factual) clauses and $C_0, C_1$ are counterfactual clauses to which the outcome $Y$ responds. Furthermore, whenever 
\begin{enumerate}[label=(\alph*)]
    \item $E_0 = E_1$, the contrast $\mathcal{C}$ is said to be counterfactual; 
    \item $C_0 = C_1$, the contrast  $\mathcal{C}$ is said to be factual.
\end{enumerate} \vspace{-0.3in}
\end{definition}
For instance, the contrast $(C_0 = \{x_0\}, C_1 = \{x_1\}, E_0 = \emptyset, E_1=\emptyset)$ corresponds to the \textit{average treatment effect (ATE)} $\ex[y_{x_1}-y_{x_0}]$. Similarly, the contrast $(C_0 = \{x_0\}, C_1 = \{x_1\}, E_0 = \{x_0\}, E_1=\{x_0\})$ corresponds to the \textit{effect of treatment on the treated (ETT)} $\ex[y_{x_1}-y_{x_0} \mid x_0]$. Many other important causal quantities can be represented as contrasts, as exemplified later on.
\begin{proposition}[Structural Basis Expansion \citep{plevcko2024causal}] \label{prop:sbe}
Counterfactual and factual contrasts admit the following structural basis expansions, respectively:
\begin{enumerate}[label=(\alph*)]
    \item Counterfactual contrast ($\mathcal{C}_{\text{ctf}}$), where $E_0 = E_1 = E$, can be expanded as
        \begin{align} 
        \ex (Y_{C_1} \mid E) - \ex(Y_{C_0} \mid E) = \sum_u \, &\big( \underbrace{Y_{C_1}(u) - Y_{C_0}(u)}_{\text{unit-level difference}} \big)
        \times \underbrace{\pr(u \mid E)}_{\text{posterior}}, \label{eq:down}
        \end{align}
    \item Factual contrast ($\mathcal{C}_{\text{factual}}$), where $C_0 = C_1 = C$, can be expanded as
        \begin{align} \label{eq:up}
             \ex(Y_{C} \mid E_1) - \ex(Y_{C} \mid E_0) = \sum_u \underbrace{Y_{C}(u)}_{\text{unit outcome}} \times \quad \big(\underbrace{\pr(u \mid E_1) - \pr(u \mid E_0)}_{\text{posterior difference}}\big).
        \end{align}
\end{enumerate}
\end{proposition}
The event $E$ in Eq.~\ref{eq:down} can be used for controlling the granularity of the population for which the effect is quantified. For instance, the choice of $C_0 = x_0, C_1 = x_1$, and $E = \emptyset$ which yields the ATE (as discussed above) is simply an average of the unit-level difference $Y_{x_1}(u) - Y_{x_0}(u)$ over all units $u$ of the population
\begin{align}
    \text{ATE}_{x_0, x_1}(y) = \sum_u (Y_{x_1}(u) - Y_{x_0}(u)) \pr(u).
\end{align}
The ETT, which is obtained with the same $C_0, C_1$ but with $E= \{X = x\}$, is a more granular notion of a total effect: 
\begin{align}
    \text{ETT}_{x_0, x_1}(y \mid x) = \sum_u (Y_{x_1}(u) - Y_{x_0}(u)) \pr(u \mid x).
\end{align}
One is still averaging the unit-level difference $Y_{x_1}(u) - Y_{x_0}(u)$ but over the subset of units that have $X(u) = x$. The weight given to each unit is given by $\pr(u \mid X = x)$, which corresponds to the probability mass of the unit $U=u$ in the event $X = x$. Further, even more granular quantifications of the effects are possible, through conditioning on $E = \{X = x, Z = z\}$, or even all the observables, $E = \{X = x, Z = z, W = w, Y = y\}$. At each level of granularity, a more localized notion of total effect is attained. It is important to note that controlling the granularity of the population works in the same for quantifying direct and indirect effects, for which different choices of $C_0, C_1$ are used (in particular, for the direct effect one can use $C_0 = \{x_0\}, C_1 = \{x_1, W_{x_0} \}$).

The notion of a spurious effect behaves somewhat differently, since the decomposition in Eq.~\ref{eq:up} shows that variations are introduced by a difference in the posterior distribution over the population. In this way, for instance, one may consider the $x$-specific spurious effect, written $x$-SE$_{x_0, x_1}(y)$, and defined as
\begin{align}
    \ex [ Y_{x_0} \mid x_1 ] - \ex [ Y_{x_0} \mid x_0] = \sum_u Y_{x_0}(u) \left( \pr(u \mid x_1) - \pr(u \mid x_0) \right).
\end{align}
The quantity measures the change in $Y$ resulting from a change conditioning on $X=x_0$ to $X=x_1$, while setting $X = x_0$ along all causal pathways. 

Importantly, the previous work considers only first-order contrasts \citep{plevcko2024causal}, and in this paper, we generalize this notion to include higher-order contrasts which are particularly useful for quantifying interaction effects. Furthermore, in Sec.~\ref{sec:population-axis}, we investigate how interactions can be analyzed at different levels of granularity of the population.
\section{Interaction Analysis} \label{sec:ia-analysis}
\begin{figure}[t]
    \centering
    \begin{subfigure}[t]{0.6\textwidth}
        \centering
        \scalebox{0.85}{ 
            \begin{tikzpicture}[
  every node/.style={font=\small, draw, rounded corners, fill=teal!30, align=center},
  mylabel/.style={
    draw=darkgreen, fill=gray!50, circle, inner sep=0pt, minimum width=0.5cm, minimum height=0.5cm
  },
  ->, >=Stealth
]
\definecolor{darkgreen}{rgb}{0.0, 0.5, 0.0}

\node[minimum width=2cm, label={[mylabel, xshift=0.2cm]north west:a}] (STR) {Structural\\ Interactions};

\node[below=1.6cm of STR, label={[mylabel, xshift=0.2cm]north west:b}] (hi-ord) {Higher Order\\ Contrasts};
\node[right=1cm of hi-ord, align=center, label={[mylabel, xshift=0.2cm]north west:c}] (Xspecific) {$x$-specific interaction\\ measures};
\node[right=1.6cm of STR, align=center, label={[mylabel, xshift=0.2cm]north west:d}] (Admissibility) {Admissibility};
\node[right=1.5cm of Admissibility, align=center, label={[mylabel, xshift=0.2cm]north west:f}] (IAtesting) {IA testing};
\node[below right=1cm and -0.6cm of IAtesting, label={[mylabel, xshift=0.2cm]north west:e}] (TV) {TV decomposition};

\draw[->, thick] (hi-ord) -- (Xspecific);
\draw[->, thick] (STR) -- (Xspecific);
\draw[->, thick] (Xspecific) -- (TV);
\draw[->, thick] (Admissibility) -- (IAtesting);
\draw[->, thick] (STR) -- (Admissibility);
\draw[->, thick] (Xspecific) -- (Admissibility);
\draw[->, thick] (Xspecific) -- (IAtesting);
\draw[->, thick] (IAtesting) -- (TV);
\end{tikzpicture}
        }
        \caption{Overview of Sec.~\ref{sec:ia-analysis}.}
        \label{fig:overview}
    \end{subfigure}%
    \hfill
    \begin{subfigure}[t]{0.33\textwidth}
        \centering
        \scalebox{1.05}{ 
            \begin{tikzpicture}
             [>=stealth, rv/.style={thick}, rvc/.style={triangle, draw, thick, minimum size=7mm}, node distance=18mm]
             \pgfsetarrows{latex-latex};
             \begin{scope}
                \node[rv] (0) at (0,1.2) {$Z$};
                \node[rv] (1) at (-1.5,0) {$X$};
                \node[rv] (2) at (0,-1.2) {$W$};
                \node[rv] (3) at (1.5,0) {$Y$};
                \draw[->] (1) -- (2);
                \draw[->] (0) -- (3);
                \draw[->] (0) -- (1);
                \path[->] (1) edge[bend left = 0] (3);
                \draw[->] (2) -- (3);
                \draw[->] (0) -- (2);
             \end{scope}
             \end{tikzpicture}
        }
        \caption{Causal diagram used in Sec.~\ref{sec:ia-analysis}.}
        \label{fig:sfm}
    \end{subfigure}
    \caption{(a) Overview of the building blocks appearing in Sec.~\ref{sec:ia-analysis}; (b) Markovian causal diagram with mediators $W$ and confounders $Z$ used in the paper.}
    \label{fig:combined}
\end{figure}
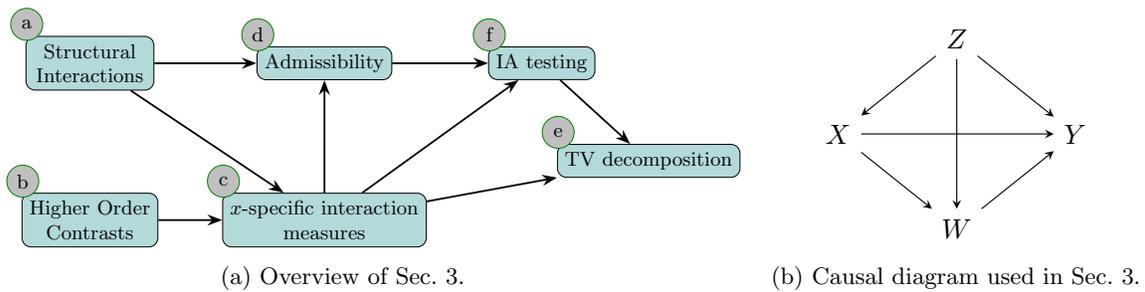

In this section, we introduce the key concepts of interaction analysis.
In Fig.~\ref{fig:overview}, we provide an overview of the building blocks appearing in this section.
Throughout, we assume that the data we are analyzing is compatible with the causal diagram in Fig.~\ref{fig:sfm} \footnote{A variation of this graph appears in previous literature, and was termed the \textit{standard fairness model (SFM)} in the context of fair machine learning \citep{plevcko2024causal}.}. The diagram consists of a treatment variable $X$ (assumed to be binary for most of the exposition), a set of confounders $Z$, a set of mediators $W$, and the outcome variable $Y$. 
Here, $Z$ and $W$ can be thought of as multi-dimensional, while $X, Y$ are singletons. We begin by explaining the overview figure and how the section is organized, and then delve into the technical details.

The notion of structural interactions (part a of Fig.~\ref{fig:overview}, Def.~\ref{def:str-ia}) is introduced first -- a definition of the ground truth for how interactions of causal pathways can be conceptualized through the lens of structural causal models. This notion would serve as a basis for one to ascertain whether pathways interact, if the knowledge of the true underlying SCM $\mathcal{M}$ was available. We then move onto defining higher-order contrasts (part b, Def.~\ref{def:higher-contrasts}), which extend the notion of a contrast considered in the previous literature \citep{plevcko2024causal}. Using the semantics of higher-order contrasts, we derive a set of causal measures intended to quantify interactions of pathways (which can be computed from data under suitable causal assumptions), which we call the $x$-specific interaction measures (part c, Def.~\ref{def:xspecific}). Then, we introduce the concept of admissibility (part d, Prop.~\ref{prop:str-admissibility}) -- which formally verifies whether a causal measure can be used for assessing the structural notion of interaction. We demonstrate that the introduced $x$-specific measures are indeed admissible with respect to the structural interaction notions, and can therefore be used for detecting structural interactions. 
After this, we demonstrate how the $x$-specific measures decompose the TV measure (part e, Thm.~\ref{thm:1st-tv-decomp}), and we further introduce the concept of interaction testing (Def.~\ref{def:ia-test}) -- testing a hypothesis that a causal measure of interaction equals $0$. If the hypothesis is not rejected, it implies that the TV measure decomposition can be made more parsimonious (part f, Alg.~\ref{algo:ia-testing}), which makes the interpretation of the decomposition easier for the data analyst. We next discuss the building blocks in Fig.~\ref{fig:overview} in the indicated order.

\subsection{Structural Interactions} \label{sec:str-ia}
To begin, we provide a structural account of pathway interactions. The structural interaction criteria we introduce can be evaluated based on the full knowledge of the structural causal model, and represent the ``ground truth'' that determines if pathways interact: 
\begin{definition}[Structural Interaction Criteria] \label{def:str-ia}
    Consider the causal diagram in Fig.~\ref{fig:conf-graph}, and let $f_y(X, Z, U_y)$ be the structural mechanism of the $Y$ variable. We say that there is no structural interaction of total and spurious effects, written Str-TE-SE = 0, if either of the following hold:
    \begin{enumerate}[label=(\roman*)]
        \item \label{cond:no-f-xz} we can write the mechanism $f_y(X, Z, U_y)$ as
        \begin{align}
        f_y(X, Z, U_y) = f^{(1)}_y(X, U_y) + f^{(2)}_y(Z, U_y).
        \end{align}
        \item \label{cond:z-to-x} $Z$ is not an input to the mechanism $f_x$ of $X$.
    \end{enumerate}
    
    Further, consider Fig.~\ref{fig:sfm} and let $f_y(X, Z, W, U_y)$ be the structural mechanism of the $Y$ variable. We say that there is no structural interaction of direct and indirect effects in the causal diagram in Fig.~\ref{fig:sfm}, written Str-DE-IE = 0, if either of the following hold:
    \begin{enumerate}[label=(\roman*)]
        \addtocounter{enumi}{2}
        \item \label{cond:no-f-xw} we can write the mechanism $f_y(X, Z, W, U_y)$ as
    \begin{align}
        f_y(X, Z, W, U_y) = f^{(1)}_y(X, Z, U_y) + f^{(2)}_y(Z, W, U_y).
    \end{align}
    \item \label{cond:x-to-w} $X$ is not an input to the mechanism $f_w$ of $W$.
    \end{enumerate}
    If there are interactions, we say that the corresponding structural criterion is equal to $1$.
\end{definition}
The definition of Str-TE-SE = 0, requires that there is no explicit functional term within $f_y$ depending on both $X, Z$ (or more generally any unobserved confounder $U_c$ with a back-door path to $X$), or that there is no back-door path between $X$ and $Y$. Similarly, the definition of Str-DE-IE requires that there is no functional term within $f_y$ depending simultaneously on both $X, W$, or that $f_w$ does not depend on the value of $X$. 
These definitions also highlight how the notions of pathway interactions, discussed in this paper, are different from \textit{mechanism interactions}. In particular, for a TE-SE interaction, having a mechanistic interaction of $X, Z$ in the $f_y$ mechanism is not sufficient\footnote{The interaction of variables in a structural mechanism $f_y$, here called ``mechanism interaction'', can be thought of as a non-parametric analogue of the concept of an \textit{interaction term}, commonly considered in parametric statistics.}; we further require that $X$ be a function of $Z$ to ascertain the interaction of pathways.
We now provide examples that illustrate the structural notions of interaction:
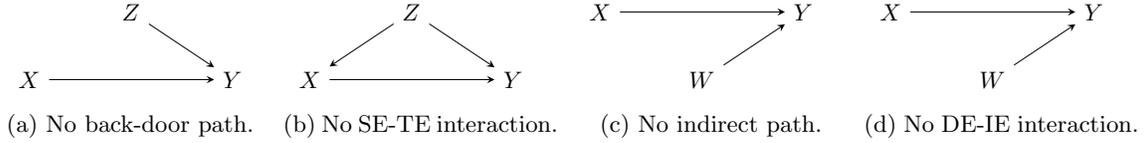
\begin{figure}
     \centering
     \begin{subfigure}[b]{0.24\textwidth}
         \centering
     \scalebox{0.9}{
     \begin{tikzpicture}
	 [>=stealth, rv/.style={thick}, rvc/.style={triangle, draw, thick, minimum size=7mm}, node distance=18mm]
	 \pgfsetarrows{latex-latex};
	 	
	 	\node[rv] (x) at (-1.5,0) {$X$};
	 	\node[rv] (z) at (0,1) {$Z$};
            \node[rv] (y) at (1.5,0) {$Y$};
	 	
	 	\draw[->] (x) -- (y);
            \draw[->] (z) -- (y);
	 \end{tikzpicture}
     }   
     \caption{No back-door path.}
     \label{fig:no-bd}
     \end{subfigure}
     \hfill
     \begin{subfigure}[b]{0.24\textwidth}
    \scalebox{0.9}{
     \begin{tikzpicture}
	 [>=stealth, rv/.style={thick}, rvc/.style={triangle, draw, thick, minimum size=7mm}, node distance=18mm]
	 \pgfsetarrows{latex-latex};
	 	
	 	\node[rv] (x) at (-1.5,0) {$X$};
	 	\node[rv] (z) at (0,1) {$Z$};
            \node[rv] (y) at (1.5,0) {$Y$};
	 	
	 	\draw[->] (x) -- (y);
            \draw[->] (z) -- (y);
            \draw[->] (z) -- (x);
	 \end{tikzpicture}
     } 
    \caption{No SE-TE interaction.}
    \label{fig:no-se-te}
     \end{subfigure}
     \hfill
     \begin{subfigure}[b]{0.24\textwidth}
    \scalebox{0.9}{
     \begin{tikzpicture}
	 [>=stealth, rv/.style={thick}, rvc/.style={triangle, draw, thick, minimum size=7mm}, node distance=18mm]
	 \pgfsetarrows{latex-latex};
	 	
	 	\node[rv] (x) at (-1.5,0) {$X$};
	 	\node[rv] (w) at (0,-1) {$W$};
            \node[rv] (y) at (1.5,0) {$Y$};
	 	
	 	\draw[->] (x) -- (y);
            \draw[->] (w) -- (y);
	 \end{tikzpicture}
     } 
    \caption{No indirect path.}
    \label{fig:no-indir}
     \end{subfigure}
     \hfill
     \begin{subfigure}[b]{0.24\textwidth}
    \scalebox{0.9}{
     \begin{tikzpicture}
	 [>=stealth, rv/.style={thick}, rvc/.style={triangle, draw, thick, minimum size=7mm}, node distance=18mm]
	 \pgfsetarrows{latex-latex};
	 	
	 	\node[rv] (x) at (-1.5,0) {$X$};
	 	\node[rv] (w) at (0,-1) {$W$};
            \node[rv] (y) at (1.5,0) {$Y$};
	 	
	 	\draw[->] (x) -- (y);
            \draw[->] (w) -- (y);
	 \end{tikzpicture}
     } 
    \caption{No DE-IE interaction.}
    \label{fig:no-de-ie}
     \end{subfigure}
     \caption{Causal diagrams used in Ex.~\ref{ex:str-ia-crit}.}
\end{figure}
\begin{example}[Structural Criteria for Interactions] \label{ex:str-ia-crit}
    Consider the SCMs $\mathcal{M}_1, \mathcal{M}_2$ given by:

\begin{minipage}{.45\linewidth}
\begin{empheq}[left={\mathcal{M}_1 = \empheqlbrace}]{align} 
  Z &\gets \text{Bernoulli}(0.5) \label{eq:m1-z}\\
  X &\gets \text{Bernoulli}(0.5) \label{eq:m1-x}\\
  Y &\gets X + Z + XZ,\label{eq:m1-y}
\end{empheq}
\end{minipage}%
\hfill \text{, } \hfill
\begin{minipage}{.45\linewidth}
\vspace{-0.1in}
\begin{empheq}[left={\mathcal{M}_2 = \empheqlbrace}]{align}
  Z &\gets \text{Bernoulli}(0.5) \label{eq:m2-z}\\
  X &\gets \text{Bernoulli}(0.5 +0.1Z) \label{eq:m2-x}\\
  Y &\gets X + Z,\label{eq:m2-y}
\end{empheq} 
\end{minipage}

\vspace{10pt}\noindent which are compatible with the diagrams in Fig.~\ref{fig:no-bd}, \ref{fig:no-se-te}, respectively. In $\mathcal{M}_1$, there is no back-door path between $X$ and $Y$, and hence there is no interaction between spurious and total effects, despite the term $XZ$ in the $f_y$ mechanism in Eq.~\ref{eq:m1-y}. In $\mathcal{M}_2$, even though a back-door path between $X$ and $Y$ exists, there is no interaction between spurious and total effects since the $f_y$ mechanism in Eq.~\ref{eq:m2-y} does not have a term involving both $X$ and $Z$. However, if there was an additional term $XZ$ in Eq.~\ref{eq:m2-y}, then the structural interaction criterion Str-TE-SE would evaluate to 1.

Consider further SCMs $\mathcal{M}_3, \mathcal{M}_4$ given by:

\begin{minipage}{.45\linewidth}
\begin{empheq}[left={\mathcal{M}_3 = \empheqlbrace}]{align} 
  X &\gets \text{Bernoulli}(0.5) \label{eq:m3-x}\\
  W &\gets \text{Bernoulli}(0.5) \label{eq:m3-w}\\
  Y &\gets X + W + XW,\label{eq:m3-y}
\end{empheq}
\end{minipage}%
\hfill \text{, } \hfill
\begin{minipage}{.45\linewidth}
\vspace{-0.1in}
\begin{empheq}[left={\mathcal{M}_4 = \empheqlbrace}]{align}
  X &\gets \text{Bernoulli}(0.5) \label{eq:m4-x}\\
  W &\gets \text{Bernoulli}(0.5 + 0.1X) \label{eq:m4-w}\\
  Y &\gets X + W,\label{eq:m4-y}
\end{empheq} 
\end{minipage}

\vspace{10pt}\noindent which are compatible with the diagrams in Fig.~\ref{fig:no-indir}, \ref{fig:no-de-ie}, respectively. In $\mathcal{M}_3$, there is no indirect path $X \to W \to Y$, since $f_w$ in Eq.~\ref{eq:m3-w} does not take $X$ as input. Therefore, regardless of the fact that $f_y$  in Eq.~\ref{eq:m3-y} contains a term $XW$, we still say there is no interaction of direct and indirect paths, written Str-DE-IE = 0. In $\mathcal{M}_4$, there exists an indirect path $X \to W \to Y$ ($f_w$ in Eq.~\ref{eq:m4-w} takes $X$ as an input), but the $f_y$ in Eq.~\ref{eq:m4-y} mechanism does not contain a term involving both $X$, and $W$, and hence again Str-DE-IE = 0. However, if there was an additional $XW$ in $f_y$ in Eq.~\ref{eq:m4-y}, then the structural interaction criterion Str-DE-IE would evaluate to 1.
\end{example}

\begin{remark}[Binary Outcomes] \label{rem:binary-y}
    The notions of interaction on the structural level introduced in Def.~\ref{def:str-ia} are presented for continuous outcomes. It is also important to extend these notions to the setting of a binary outcome $Y$.
The key idea in the binary case is to replace the structural mechanism $f_y$ with its probability counterpart, defined by:
\begin{align}
    p_y(x, z, w) := \ex_{u_y}[Y_{x, z, w} ] = \pr_{U_y}(Y_{x, z, w} = 1),
\end{align}
The structural mechanism $f_y$ returns a binary value of $Y$, while $p_y$ returns the probability of $Y$ belonging to class $1$ given covariates $X=x, Z=z, W=w$. When evaluating the structural notions for a binary outcome, we use $p_y$ in place of $f_y$ throughout Def.~\ref{def:str-ia}.
\end{remark}
Furthermore, for binary outcomes, it may also be relevant to investigate interactions on the log-risk or log-odds scales. These extensions are discussed in Appendix~\ref{appendix:lr-lo-scales}. 

\subsection{Higher-Order Contrasts \& x-specific Measures}
We next introduce the distinction between first order, second order, and third order contrasts, and then instantiate some well-known first and second-order contrasts:
\begin{definition}[First, Second, and Third Order Contrasts] \label{def:higher-contrasts}
    Let $C_0, C_1, C, C'$ be counterfactual clauses, and $E_0, E_1$ be observed events. We categorize the following types of contrasts:
    \begin{enumerate}[label=(\arabic*)]
        \item A first order causal contrast $\mathcal{C}^1(C_0, C_1, E, E)$, of the form
        \begin{align}
            \sum_u \big[Y_{C_1}(u) - Y_{C_0}(u)\big] \times \pr(u \mid E).
        \end{align}
        \item A first order spurious contrast $\mathcal{C}^1(C, C, E_0, E_1)$, of the form
        \begin{align}
            \sum_u Y_{C}(u) \times \big[ \pr(u \mid E_1) - \pr(u \mid E_0) \big].
        \end{align}
        \item A second order causal-causal contrast $\mathcal{C}^2(C_0, C_1, C, C', E, E)$, of the form
        \begin{align}
            \sum_u \big[(Y_{C_1, C'}(u) - Y_{C_0, C'}(u)) - (Y_{C_1, C}(u) - Y_{C_0, C}(u))\big] \times \pr(u \mid E).
        \end{align}
        \item A second order causal-spurious contrast $\mathcal{C}^2(C_0, C_1, E_0, E_1)$, of the form
        \begin{align}
            \sum_u \big[Y_{C_1}(u) - Y_{C_0}(u)\big] \times \big[ \pr(u \mid E_1) - \pr(u \mid E_0) \big].
        \end{align}
        \item A third order causal-causal-spurious contrast $\mathcal{C}^3(C_0, C_1, C, C', E_0, E_1)$, of the form
        \begin{align}
            \sum_u \big[(Y_{C_1, C}(u) - Y_{C_0, C}(u)) - (Y_{C_1, C'}(u) - Y_{C_0, C'}(u))\big] \times \big[ \pr(u \mid E_1) - \pr(u \mid E_0) \big].
        \end{align}
    \end{enumerate}
\end{definition}
\begin{example}
    The average treatment effect (ATE) is a first order causal contrast since it can be written as:
    \begin{align}
        \text{ATE}_{x_0, x_1}(y) = \sum_u \big[ Y_{x_1}(u) - Y_{x_0}(u)\big] \times \pr(u) = \mathcal{C}^1(\{x_0\}, \{x_1\}, \emptyset, \emptyset).
    \end{align}
    The measure of causal-spurious effect interaction introduced in Eq.~\ref{eq:intro-TE-SE} of Ex.~\ref{ex:intro-decompositions}, can be written as
    \begin{align}
        \ex[Y_{x_1} - Y_{x_0} \mid x_1] - \ex[Y_{x_1} - Y_{x_0} \mid x_0] &= \sum_u \big[ Y_{x_1}(u) - Y_{x_0}(u) \big] \times \big[ \pr (u \mid x_1) - \pr(u \mid x_0) \big] \\
        &= \mathcal{C}^2(\{x_0\}, \{x_1\}, \{x_0\}, \{x_1\}),
    \end{align}
    and is hence a second-order causal-spurious contrast.
\end{example}
Using the semantics of higher order contrasts, we define the following $x$-specific causal measures:
\begin{definition}[$x$-specific Effects] \label{def:xspecific}
The $x$-\{total, direct, indirect, spurious\} effects are defined as follows:
    \begin{align}
        x\text{-TE}_{x_0, x_1}(y\mid x) &= \ex(Y_{x_1} \mid x) - \ex(Y_{x_0}\mid x)  \\
        x\text{-DE}_{x_0, x_1}(y\mid x) &= \ex(Y_{x_1, W_{x_0}} \mid x) - \ex(Y_{x_0}\mid x)  \\
        x\text{-IE}_{x_0, x_1}(y\mid x) &= \ex(Y_{x_0, W_{x_1}}\mid x) - \ex(Y_{x_0} \mid x)\\
        x\text{-SE}_{x_0, x_1}(y) &= \ex(Y_{x_0} \mid x_1) - \ex(Y_{x_0} \mid x_0). 
\end{align}
The interactive effects $x$-{direct-indirect} and $x$-{total-spurious} are defined as:
\begin{align}
    x\text{-DE-IE}_{x_0, x_1}(y \mid x) &= [\ex(Y_{x_1, W_{x_1}} \mid x) - \ex(Y_{x_0, W_{x_1}} \mid x)] - [\ex(Y_{x_1, W_{x_0}} \mid x) - \ex(Y_{x_0, W_{x_0}} \mid x)]  \\
        x\text{-TE-SE}_{x_0, x_1}(y) &= [\ex(Y_{x_1} \mid x_1) - \ex(Y_{x_0} \mid x_1)] - [\ex(Y_{x_1} \mid x_0) - \ex(Y_{x_0} \mid x_0)].
\end{align}
\end{definition}
To better understand what the measures of interaction quantify, we look at their structural basis expansion. The effect $x\text{-DE-IE}_{x_0, x_1}(y \mid x)$ can be written as:
\begin{align} \label{eq:x-de-ie-sbe}
    x\text{-DE-IE}_{x_0, x_1}(y \mid x) = \sum_u [\underbrace{ (Y_{x_1, W_{x_1}} - Y_{x_0, W_{x_1}})(u) - (Y_{x_1, W_{x_0}} - Y_{x_0, W_{x_0}})(u)}_{\text{Term }T_1} ]  \times \underbrace{\pr(u \mid x)}_{\text{Term }T_2}.
\end{align}
The unit-level difference term $T_1$ appearing in Eq.~\ref{eq:x-de-ie-sbe} equals
\begin{align}
    \underbrace{Y_{x_1, W_{x_1}}(u) - Y_{x_0, W_{x_1}}(u)}_{x_0\to x_1 \text{ DE with } W_{x_1}} - \underbrace{Y_{x_1, W_{x_0}}(u) - Y_{x_0, W_{x_0}}(u)}_{x_0\to x_1 \text{ DE with } W_{x_0}},
\end{align}
and can thus be understood as how much changing $W_{x_0}$ to $W_{x_1}$ modifies the size of the direct effect of the transition $x_0 \to x_1$. Due to symmetry, this unit-level difference can be also be written as:
\begin{align}
    \underbrace{Y_{x_1, W_{x_1}}(u) - Y_{x_1, W_{x_0}}(u)}_{x_0\to x_1 \text{ IE with } X = {x_1}\to Y } - \underbrace{Y_{x_0, W_{x_1}}(u) - Y_{x_0, W_{x_0}}(u) }_{x_0\to x_1 \text{ IE with } X = {x_0}\to Y}.
\end{align}
In words, the interactive effect can also be understood as how much changing $x_0$ to $x_1$ along the direct effect modifies the sizes of the indirect effect of a $x_0 \to x_1$ transition. To compute the $x$-DE-IE quantity, the unit-level difference $T_1$ is averaged using the posterior distribution $\pr(u \mid x)$, which gives the posterior weight 
\begin{align}
    \frac{\pr(u)}{\sum_{u': X(u') = x} \pr(u')},
\end{align}
to each unit $u$ compatible with $X(u) = x$. In this sense, the measure in Eq.~\ref{eq:x-de-ie-sbe} is a weighted average of a second-order unit-level difference $T_1$, with the associated weight being proportional to the probability of the unit $u$ within the event $X(u) = x$. Alternatively, $x$-DE-IE can also be expressed through $x$-specific direct or indirect effects, and written as:
\begin{align}
    x\text{-DE-IE}_{x_0, x_1}(y \mid x) &= - x\text{-DE}_{x_1, x_0}(y \mid x) - x\text{-DE}_{x_0, x_1}(y \mid x) \\
    &= - x\text{-IE}_{x_1, x_0}(y \mid x) - x\text{-IE}_{x_0, x_1}(y \mid x).
\end{align}

We now turn to understanding the total-spurious interaction effect, which can be written using the structural basis expansion as:
\begin{align}
    x\text{-TE-SE}_{x_0, x_1}(y) = \sum_u \big[ \underbrace{Y_{x_1}(u) - Y_{x_0}(u)}_{\text{Term } T_3} \big] \times \big[ \underbrace{\pr(u \mid x_1) - \pr(u \mid x_0)}_{\text{Term } T_4}\big].
\end{align}
The first-order unit-level difference $Y_{x_1}(u) - Y_{x_0}(u)$ in term $T_3$ is integrated against the first-order difference in posterior weighing distributions $\pr(u \mid x_1) - \pr(u \mid x_0)$. The term $T_3$ captures a causal effect, while the difference in the posteria in term $T_4$ captures a spurious effect, by comparing how likely a unit $u$ is to be associated with the event $X(u) = x_1$ vs. $X(u) = x_0$. In this way, the $x$-TE-SE quantity captures the interaction of the total and spurious effects. Once again, by symmetry, the measure can be expressed using $x$-specific total or spurious effects, as follows:
\begin{align}
    x\text{-TE-SE}_{x_0, x_1}(y) &= x\text{-TE}_{x_0, x_1}(y \mid x_1) - x\text{-TE}_{x_0, x_1}(y \mid x_0) \\
    &= - x\text{-SE}_{x_1, x_0}(y) - x\text{-SE}_{x_0, x_1}(y).
\end{align}

\subsection{Admissibility of x-specific Measures}
So far, we introduced the notions of structural interaction (Def.~\ref{def:str-ia}) and the $x$-specific measures of interaction (Def.~\ref{def:xspecific}). The key question we ask next is whether the $x$-specific measures of interaction are useful for capturing structural interactions. As shown in the following key result, the answer is affirmative:
\begin{proposition}[Structural Admissibility of Interaction Tests] \label{prop:str-admissibility}
    \begin{align}
        \text{Str-TE-SE} = 0 &\implies x\text{-TE-SE}_{x_0, x_1}(y) = 0, \\
        \text{Str-DE-IE} = 0 &\implies x\text{-DE-IE}_{x_0, x_1}(y\mid x) = 0.
    \end{align}
\end{proposition}
In words, the proposition shows that whenever there is no total-spurious interaction at the structural level, the corresponding interaction effect $x$-TE-SE equals $0$. The same also holds for the direct-indirect interaction and the corresponding $x$-DE-IE interaction effect. The key of the proposition is in the contrapositive of the mentioned statements. That is, whenever we find that 
\begin{align}
    x\text{-TE-SE}_{x_0, x_1}(y) \neq 0,
\end{align}
it implies that an interaction at the structural level must exist. The usefulness of this observation comes from the fact that the effect $x\text{-TE-SE}_{x_0, x_1}(y)$ can be estimated non-parametrically from data, and testing the hypothesis $H_0: x\text{-TE-SE}_{x_0, x_1}(y) = 0$ provides a valid test for testing if Str-TE-SE = 0 (and analogously for the DE-IE interaction). Therefore, the admissibility result in Prop.~\ref{prop:str-admissibility} allows us to move from evaluating the existence of structural interactions based on the knowledge of the structural causal model (almost never available to us), to a practical approach which can be carried out based on data and suitable causal assumptions (such as those encoded in the diagram in Fig.~\ref{fig:sfm}).

\subsection{Decomposing the TV Measure}
Armed with the structural understanding of different (interaction) effects, we can now decompose the TV measure: 
\begin{theorem}[TV Decomposition with Interactions] \label{thm:1st-tv-decomp}
    The total variation (TV) measure can be decomposed as:
    \begin{align} \label{eq:tv-decomp-te-se}
        \text{TV}_{x_0, x_1}(y) &= {x\text{-TE}_{x_0, x_1}(y\mid x_0)} + {x\text{-SE}_{x_0, x_1}(y)} + x\text{-TE-SE}_{x_0, x_1}(y).
    \end{align}
    Furthermore, the TV measure can also be decomposed as:
    \begin{align} \label{eq:tv-decomp-de-ie-se}
        \text{TV}_{x_0, x_1}(y) &= {x\text{-DE}_{x_0, x_1}(y\mid x_0)} + {x\text{-IE}_{x_0, x_1}(y\mid x_0)} +  x\text{-DE-IE}_{x_0, x_1}(y \mid x_0)\\ 
        &\quad + {x\text{-SE}_{x_0, x_1}(y)} + x\text{-TE-SE}_{x_0, x_1}(y). \nonumber
    \end{align}
\end{theorem}
Thm.~\ref{thm:1st-tv-decomp} represents the first major result of the paper. The TV measure is decomposed to feature explicit interaction effects of the direct/indirect pathways, and also the total/spurious pathways. Furthermore, all the transitions appearing for direct, indirect, and spurious effects are $x_0 \to x_1$, and there are no subtractions necessary (i.e., all the effects are added up). This is another benefit when interpreting the decomposition. 

\subsection{Interaction Testing}
\begin{algorithm}[t]
    \caption{Interaction Testing for TV Decomposition}
     \begin{algorithmic}[1]
        \Statex \textbullet~\textbf{Inputs:} Causal Diagram $\mathcal{G}$, Observational Data $\mathcal{D}$
        \Statex Compute the estimate of the total-spurious interaction effect $x\text{-TE-SE}_{x_0, x_1}(y)$, and its 95\% confidence interval. Test the hypothesis
        \begin{align}
            H^{\text{TE-SE}}_0: x\text{-TE-SE}_{x_0, x_1}(y) = 0.
        \end{align}
        \Statex Compute the estimate of the direct-indirect interaction effect $x\text{-DE-IE}_{x_0, x_1}(y \mid x_0)$, and its 95\% confidence interval. Test the hypothesis
        \begin{align}
            H^{\text{DE-IE}}_0: x\text{-DE-IE}_{x_0, x_1}(y \mid x_0) = 0.
        \end{align}
        \Statex \begin{itemize}
            \item if neither $H^{\text{TE-SE}}_0, H^{\text{DE-IE}}_0$ are rejected, return the decomposition
            \begin{align}
                \text{TV}_{x_0, x_1}(y) &= {x\text{-DE}_{x_0, x_1}(y\mid x_0)} + {x\text{-IE}_{x_0, x_1}(y\mid x_0)} + {x\text{-SE}_{x_0, x_1}(y)} 
            \end{align}
            \item if only $H^{\text{DE-IE}}_0$ is rejected, return the decomposition
            \begin{align} \label{eq:h-te-se-not-reject}
            \text{TV}_{x_0, x_1}(y) &= {x\text{-DE}_{x_0, x_1}(y\mid x_0)} + {x\text{-IE}_{x_0, x_1}(y\mid x_0)} +  x\text{-DE-IE}_{x_0, x_1}(y \mid x_0)\\ &\quad
            + {x\text{-SE}_{x_0, x_1}(y)}. \nonumber
         \end{align}
         \item if only $H^{\text{TE-SE}}_0$ is rejected, return the decomposition
            \begin{align} \label{eq:h-de-ie-not-reject}
            \hspace{-5pt}\text{TV}_{x_0, x_1}(y) &= {x\text{-DE}_{x_0, x_1}(y\mid x_0)} + {x\text{-IE}_{x_0, x_1}(y\mid x_0)} 
            + {x\text{-SE}_{x_0, x_1}(y)} + {x\text{-TE-SE}_{x_0, x_1}(y)}. 
         \end{align}
        \item if both $H^{\text{TE-SE}}_0, H^{\text{DE-IE}}_0$ are rejected, return the TV decomposition in Eq.~\ref{eq:tv-decomp-de-ie-se}.
        \end{itemize}
        \Statex \textbullet~\textbf{Output:} TV decomposition with parsimony.
     \end{algorithmic}
     \label{algo:ia-testing}
\end{algorithm}
We next move to the aspect of variation analysis called interaction testing. Previously, measure for quantifying interactions were developed, and these have been shown as admissible with respect to the structural notions of interaction. Next, a formal procedure to test if an interaction is significant (and, implicitly, to test whether an interaction exists on the structural level) is introduced. We begin with the definition of an interaction test:
\begin{definition}[Interaction Test] \label{def:ia-test}
    Let $\mathcal{C}$ be a contrast admissible with respect to a structural interaction criterion.
    We say that a hypothesis test of the form
    \begin{align}
        H_0: \mathcal{C} (\mathcal{D}) = 0
    \end{align}
    is called an interaction test.
\end{definition}
\begin{example}[Total-Spurious Interaction Test]
    Consider the $x$-specific total-spurious interaction effect $x\text{-TE-SE}_{x_0, x_1}(y)$. The hypothesis test
    \begin{align}
        H_0 : x\text{-TE-SE}_{x_0, x_1}(y) = 0
    \end{align}
    is an interaction test for the structural criterion Str-TE-SE.
\end{example}
The reasoning behind interaction testing is rather simple. Generally, interactions make the interpretation of causal effect decompositions more difficult. In statistics, there is often a preference for parsimony, whenever such parsimony is justifiable. Therefore, we propose an approach in which an interaction effect $\mathcal{C}$ is tested against $0$, based on some available data $\mathcal{D}$. If there is no evidence of the effect being different from $0$, we may use a more parsimonious version of the TV decomposition.

Based on the above, we propose an approach for interaction testing presented in Alg.~\ref{algo:ia-testing}. 
First, the hypothesis $x\text{-TE-SE}_{x_0, x_1}(y) = 0$ is tested, and if this hypothesis is not rejected, one can use a more parsimonious representation of the TV decomposition, in which the effect $x\text{-TE-SE}_{x_0, x_1}(y) = 0$ is removed (see Eq.~\ref{eq:h-te-se-not-reject}). Second, the hypothesis $x\text{-DE-IE}_{x_0, x_1}(y \mid x_0) = 0$ is tested. If the hypothesis is not rejected, one can again use a more parsimonious TV decomposition, in which the effect $x\text{-DE-IE}_{x_0, x_1}(y \mid x_0)$ is removed (see Eq.~\ref{eq:h-de-ie-not-reject}). More parsimonious decompositions, naturally, lead to more easily interpretable decompositions for practitioners.
\section{More Granular Interactions}
So far, we described an approach for testing the interaction of total and spurious effects (TE-SE), and direct and indirect effects (DE-IE). 
The summary of the approach is shown in Tab.~\ref{tab:ia-testing-summary}. Two types of interactions tests were considered, and the implications of these tests for the TV decomposition are provided.

Often, however, the objective may be an even more granular analysis: to distinguish between direct and spurious, indirect and spurious, or even direct-indirect-spurious interactions. The following example illustrates why considering such more granular interactions is more involved than considering the TE-SE and DE-IE interactions:
\begin{example}[DE-SE Interaction] \label{ex:de-se-ia}
    Consider the following structural causal model $\mathcal{M}$:
    \begin{align}
        Z &\gets \eps_z \label{eq:ex-de-se-1}\\
        X &\gets \text{Bern}(\frac{e^Z}{1 + e^Z}) \\
        W &\gets \eps_w + Z \\
        Y &\gets W + XW^2, \label{eq:ex-de-se-4}
    \end{align}
    where $\eps_z, \eps_w \sim N(0, 1)$. The model is compatible with the causal diagram in Fig.~\ref{fig:sfm}. When computing the quantity 
    \begin{align} \label{eq:de-se-ia}
        x\text{-DE-SE}_{x_0, x_1}(y) := \ex[Y_{x_1, W_{x_0}} - Y_{x_0} \mid x_1] - \ex[Y_{x_1, W_{x_0}} - Y_{x_0} \mid x_0],
    \end{align}
    we find that it equals $1$. The quantity in Eq.~\ref{eq:de-se-ia} attempts to measure how much the direct effect of a $x_0 \to x_1$ transition, written $Y_{x_1, W_{x_0}} - Y_{x_0}$, changes for units with $X(u) = x_1$ vs. $X(u) = x_0$. Therefore, the quantity measures the interaction of direct and spurious paths, and in $\mathcal{M}$, it is different from $0$. However, when looking at the $f_y$ mechanism, we see that there is no term corresponding to the $X, Z$ interaction.
\end{example}
The reader will note the issue at hand -- in the diagram in Fig.~\ref{fig:sfm}, the spurious effect of $X$ on $Y$ may be transmitted along two different paths: firstly, it may be transmitted along the $X \gets Z \to Y$ path, but may also be transmitted along the $X \gets Z \to W \to Y$ path. It is the latter path that is active in the SCM $\mathcal{M}$ in the above example. Previously, when testing the TE-SE interaction in the diagram in Fig.~\ref{fig:conf-graph}, there was a single causal path $X \to Y$, and a single spurious path $X \gets Z \to Y$. Similarly, for the direct and indirect paths in Fig.~\ref{fig:med-graph} (or Fig.~\ref{fig:sfm}), there was a single direct $X \to Y$ path and a single indirect $X \to W \to Y$ path. This one-to-one correspondence between an effect (total/spurious or direct/indirect) and a causal path entering $Y$ implied that the existence of an interaction can be easily determined at the structural level, by inspecting the $f_y$ mechanism (see Def.~\ref{def:str-ia}) and checking for either the existence of a back-door path between $X$ and $Y$ (in case of TE-SE interaction) or an indirect path (in case of DE-IE interaction). In Ex.~\ref{ex:de-se-ia}, since there are two spurious paths from $X$ to $Y$, more care needs to be taken. Clearly, an interaction of spurious and direct paths, for instance, may also depend on the functional inputs of the $f_w$ mechanism. 
\begin{table}[t]
\centering
\scalebox{0.65}{
\renewcommand{\arraystretch}{2}
\begin{tabular}{|c|c|c|c|c|}
\hline
Interaction & Structural Test & Diagram & Interaction Test & Implication for Decomposition \\[1mm]
\hline
TE $\otimes$ SE & \makecell{no $f_y(x,z,u_y)$ term in $f_y$ \\ OR \\ no backdoor $(X, Y)$ path}
 & Fig.~\ref{fig:conf-graph} & $x\text{-SE}_{x_0, x_1}(y) = -x\text{-SE}_{x_1, x_0}(y)$ & $\text{TV}_{x_0, x_1}(y) = {x\text{-TE}_{x_0, x_1}(y\mid x_0)} + {x\text{-SE}_{x_0, x_1}(y)}$ \\[5mm]
\hline
DE $\otimes$ IE & \makecell{no $f_y(x,w,u_y)$ term in $f_y$ \\ OR \\ no $X\to W \to Y$ path} & Fig.~\ref{fig:med-graph} & $x\text{-DE}_{x_0, x_1}(y \mid x) = -x\text{-DE}_{x_1, x_0}(y \mid x)$  & 
${x\text{-TE}_{x_0, x_1}(y\mid x_0)} = {x\text{-DE}_{x_0, x_1}(y\mid x_0)} + {x\text{-IE}_{x_0, x_1}(y\mid x_0)}$\\[4mm]
\hline
\end{tabular}
}
\caption{Summary table of interaction tests and their implications.}
\label{tab:ia-testing-summary}
\end{table}

Therefore, understanding the interactions of direct/spurious, indirect/spurious, or even the direct/indirect/spurious effects will require a slightly more involved approach than for total/spurious and direct/indirect interactions. We begin by defining the more granular interactions at the structural level:
\begin{definition}[Granular Structural Interaction Criteria] \label{def:str-ia-granular}
    Consider the causal diagram in Fig.~\ref{fig:sfm}, and let $f_y(X, Z, W, U_y)$ be the structural mechanism of the $Y$ variable. We say that there is no structural interaction of direct and spurious effects, written Str-DE-SE = 0, if either of the following hold:
    \begin{enumerate}[label=(\roman*)]
        \item $Z$ is not an argument of $f_x$,
        \item $Z$ is not an argument of $f_w$ and we can write the mechanism $f_y(X, Z, W, U_y)$ as
        \begin{align}
        f_y(X, Z, W, U_y) = f^{(1)}_y(X, W, U_y) + f^{(2)}_y(Z, W, U_y),
        \end{align}
        \item we can write the mechanism $f_y(X, Z, W, U_y)$ as
        \begin{align}
        f_y(X, Z, W, U_y) = f^{(1)}_y(X, U_y) + f^{(2)}_y(Z, W, U_y).
        \end{align}
    \end{enumerate}
    
    Further, we say that there is no structural interaction of indirect and spurious effects, written Str-IE-SE = 0, if either of the following hold:
    \begin{enumerate}[label=(\roman*)]
        \item $Z$ is not an argument of $f_x$,
        \item $X$ is not an argument of $f_w$,
        \item we can write the mechanism $f_w$ as
        \begin{align}
            f_w(X, Z, U_w) = f^{(1)}_w(X, U_w) + f^{(2)}_w(Z, U_w),
        \end{align}
        and the mechanism $f_y$ as
        \begin{align}
        f_y(X, Z, W, U_y) = f^{(1)}_y(X, Z, U_y) + W f^{(2)}_y(X, U_y),
        \end{align}
        \item $W$ is not an argument of $f_y$.
    \end{enumerate}

    We say that there is no structural interaction of direct, indirect, and spurious effects, written Str-DE-IE-SE = 0, if either of the following hold:
    \begin{enumerate}[label=(\roman*)]
        \item $Z$ is not an argument of $f_x$,
        \item $X$ is not an argument of $f_w$,
        \item we can write the mechanism $f_w$ as
        \begin{align}
            f_w(X, Z, U_w) = f^{(1)}_w(X, U_y) + f^{(2)}_w(Z, U_y),
        \end{align}
        and the mechanism $f_y$ as
        \begin{align}
        f_y(X, Z, W, U_y) = f^{(1)}_y(X, Z, U_y) + W f^{(2)}_y(X, U_y) + f^{(3)}_y(Z, W, U_y),
        \end{align}
        \item the mechanism $f_y$ can be written as
        \begin{align}
        f_y(X, Z, W, U_y) = f^{(1)}_y(X, Z, U_y) + f^{(2)}_y(Z, W, U_y).
        \end{align}
    \end{enumerate}
    If there are interactions, we say that the corresponding structural criterion is equal to $1$.
\end{definition}
In Appendix~\ref{appendix:str-decision-trees}, we provide a further discussion on how to arrive at the above definitions.
We can now define the quantities that measure the interactions of the above-introduced effects. 
\begin{definition}[Granular Effect Interactions] \label{def:xspecific-granular}
    Consider the following effect interactions:
    \begin{align} \label{eq:x-de-se}
        x\text{-DE-SE}_{x_0, x_1}(y) &= [\ex(Y_{x_1, W_{x_0}} \mid x_1) - \ex(Y_{x_0} \mid x_1)] - [\ex(Y_{x_1, W_{x_0}} \mid x_0) - \ex(Y_{x_0} \mid x_0)] \\ \label{eq:x-ie-se}
        x\text{-IE-SE}_{x_0, x_1}(y) &= [\ex(Y_{x_0, W_{x_1}} \mid x_1) - \ex(Y_{x_0} \mid x_1)] - [\ex(Y_{x_0, W_{x_1}} \mid x_0) - \ex(Y_{x_0} \mid x_0)] \\
        x\text{-DE-IE-SE}_{x_0, x_1}(y) &= [\ex(Y_{x_1} \mid x_1) - \ex(Y_{x_0, W_{x_1}} \mid x_1)] - [\ex(Y_{x_1, W_{x_0}} \mid x_1) - \ex(Y_{x_0} \mid x_1)] \\ \label{eq:x-de-ie-se}
        &\quad - \big[ [\ex(Y_{x_1} \mid x_0) - \ex(Y_{x_0, W_{x_1}} \mid x_0)] - [\ex(Y_{x_1, W_{x_0}} \mid x_0) - \ex(Y_{x_0} \mid x_0)]   \big]
    \end{align}
\end{definition}
Again, to obtain a better insight into the quantities, we look at their structural basis expansion. For instance, the quantity $x$-DE-SE can be expanded as:
\begin{align}
    x\text{-DE-SE}_{x_0, x_1}(y) = \sum_u [Y_{x_1, W_{x_0}}(u) - Y_{x_0}(u)] \times [\pr(u \mid x_1) - \pr(u \mid x_0)].
\end{align}
The unit-level direct effect, $Y_{x_1, W_{x_0}}(u) - Y_{x_0}(u)$, is integrated against a posterior difference $\pr(u \mid x_1) - \pr(u \mid x_0)$ that induces a difference along the spurious path. In this way, we see that the quantity is a second-order causal-spurious contrast (recall Def.~\ref{def:higher-contrasts}), measuring the interaction of the direct and spurious effects. For indirect-spurious interactions, we have the quantity
\begin{align}
    x\text{-IE-SE}_{x_0, x_1}(y) = \sum_u [Y_{x_0, W_{x_1}}(u) - Y_{x_0}(u)] \times [\pr(u \mid x_1) - \pr(u \mid x_0)].
\end{align}
The interpretation is again very similar, where the unit-level indirect effect $Y_{x_0, W_{x_1}}(u) - Y_{x_0}(u)$ is integrated against the posterior difference $\pr(u \mid x_1) - \pr(u \mid x_0)$. The quantity, therefore, captures the indirect-spurious interaction. Finally, for quantifying the three-way interaction, we can use the quantity
\begin{align}
    x\text{-DE-IE-SE}_{x_0, x_1}(y) = \sum_u & [ (Y_{x_1, W_{x_1}} - Y_{x_0, W_{x_1}})(u) - (Y_{x_1, W_{x_0}} - Y_{x_0, W_{x_0}})(u) ] \\&
    \times [\pr(u \mid x_1) - \pr(u \mid x_0)].
\end{align}
The quantity is therefore a third-order causal-causal-spurious contrast. The unit-level interaction effect $(Y_{x_1, W_{x_0}} - Y_{x_0, W_{x_0}})(u) - (Y_{x_1, W_{x_1}} - Y_{x_0, W_{x_1}})(u)$ is integrated against the posterior difference $\pr(u \mid x_1) - \pr(u \mid x_0)$. In this way, one can quantify the interaction of the direct-indirect interaction and the spurious effect, effectively capturing a three-way interaction effect. After understanding the granular interaction effects, we next prove an extended decomposition of the TV measure:
\begin{theorem}[Total TV Decomposition] \label{thm:2nd-tv-decomp}
    The TV measure admits the following decomposition:
    \begin{align} \label{eq:tv-first-order}
        \text{TV}_{x_0, x_1}(y) &= {x\text{-DE}_{x_0, x_1}(y\mid x_0)} + {x\text{-IE}_{x_0, x_1}(y\mid x_0)} + {x\text{-SE}_{x_0, x_1}(y)} \\ \label{eq:tv-second-order}
        &\quad + x\text{-DE-IE}_{x_0, x_1}(y \mid x_0) + x\text{-DE-SE}_{x_0, x_1}(y) + x\text{-IE-SE}_{x_0, x_1}(y)\\ \label{eq:tv-third-order}
        &\quad  + x\text{-DE-IE-SE}_{x_0, x_1}(y). 
    \end{align}
\end{theorem}
The above theorem is a crown result of the paper. The TV measure can be decomposed into first order direct, indirect, and spurious effects (Line~\ref{eq:tv-first-order}), the direct-indirect, direct-spurious, indirect-spurious interaction effects (Line~\ref{eq:tv-second-order}), and the three-way interaction of direct, indirect, and spurious effects (Line~\ref{eq:tv-third-order}). Once again, one may use an interaction testing procedure as in Alg.~\ref{algo:ia-testing} to obtain a more parsimonious TV representation in case some of the second or third order effects are not significantly different from $0$. Once again, this type of interaction testing is justified through the following admissibility result (analogous to Prop.~\ref{prop:str-admissibility}):
\begin{proposition}[Structural Admissibility of Granular Interaction Tests] \label{prop:str-admissibility-granular}
    \begin{align}
        \text{Str-DE-SE} = 0 &\implies x\text{-DE-SE}_{x_0, x_1}(y) = 0, \\
        \text{Str-IE-SE} = 0 &\implies x\text{-IE-SE}_{x_0, x_1}(y) = 0, \\
        \text{Str-DE-IE-SE} = 0 &\implies x\text{-DE-IE-SE}_{x_0, x_1}(y) = 0.
    \end{align}
\end{proposition}

\paragraph{Practical Implications.} There are a number of practical implications resulting from the TV decomposition in Eqs.~\ref{eq:tv-first-order}-\ref{eq:tv-third-order}. Firstly, when the hypothesis
\begin{align}
    H^{\text{DE-IE}}_0: x\text{-DE-IE}_{x_0, x_1}(y \mid x_0) = 0
\end{align}
is rejected, the practitioner can conclude that the mediator value $W=w$ modifies the direct effect, namely that the controlled direct effect 
\begin{align}
    \text{CDE}_{x_0, x_1}(y_{w}) = E[Y_{x_1, w} - Y_{x_0, w}]
\end{align}
varies according to levels of $W = w$. Investigating such an effect along different $W = w$ levels may therefore be informative. Similarly, the existence of DE-SE or IE-SE interactions implies that there is heterogeneity in the controlled direct effect 
\begin{align}
    \text{CDE}_{x_0, x_1}(y_{w, z}) = E[Y_{x_1, z, w} - Y_{x_0, z, w}],
\end{align}
meaning that the practitioner may wish to focus on the heterogeneity of the $\text{CDE}_{x_0, x_1}(y_{w, z})$ according to levels of $Z = z, W=w$ to learn more about the phenomenon under study. An alternative view of the DE-SE interaction is that the DE may vary according to $Z =z$, namely that
\begin{align}
    z\text{-DE}_{x_0, x_1}(y \mid z) = \ex[Y_{x_1, W_{x_0}} - Y_{x_0} \mid z] 
\end{align}
may vary with $Z = z$ (and similarly for $Z = z$). We discuss some tools for investigating this in the next section.

\subsection{Population Granularity} \label{sec:population-axis}
In this section, we discuss population granularity, and the testing of interactions along different subpopulations of the data. We begin with a motivating example:
\begin{example}[DE-IE Interaction Testing] \label{ex:de-ie-granularity}
    Consider the SCM $\mathcal{M}$ given by:
    \begin{align}
        Z &\gets \text{Bernoulli}(0.5) \label{eq:scm-de-ie-power-z}\\
        X &\gets \text{Bernoulli}(0.5) \\
        W &\gets 1-X \\
        Y &\gets (2Z -1)XW. \label{eq:scm-de-ie-power-y}
    \end{align}
    If we test for the interaction of direct and indirect effects, using the measure 
    \begin{align}
        x\text{-DE-IE}_{x_0, x_1}(y \mid x) &= [\ex(Y_{x_1, W_{x_0}} \mid x) - \ex(Y_{x_0, W_{x_0}} \mid x)] - [\ex(Y_{x_1, W_{x_1}} \mid x) - \ex(Y_{x_0, W_{x_1}} \mid x)]
    \end{align}
    we find that for both $x \in \{ x_0, x_1 \}$, we have
    \begin{align}
        x\text{-DE-IE}_{x_0, x_1}(y \mid x) = 0.
    \end{align}
    Therefore, based on the measure $x\text{-DE-IE}_{x_0, x_1}(y \mid x)$ we cannot detect an interaction of direct and indirect effects, even though $X$ influences $W$ and there is a term with an interaction $XW$ in $f_y$. 

    However, when we consider the difference $[Y_{x_1, W_{x_0}} - Y_{x_0, W_{x_0}}](u) - [Y_{x_1, W_{x_1}} - Y_{x_0, W_{x_1}}](u)$ averaged across units $u$ with $Z(u) = 0$ and $Z(u) = 1$, respectively, we find that
    \begin{align}
        [\ex(Y_{x_1, W_{x_0}} \mid z_1) - \ex(Y_{x_0, W_{x_0}} \mid z_1)] - [\ex(Y_{x_1, W_{x_1}} \mid z_1) - \ex(Y_{x_0, W_{x_1}} \mid z_1)] &= 1 \\
        [\ex(Y_{x_1, W_{x_0}} \mid z_0) - \ex(Y_{x_0, W_{x_0}} \mid z_0)] - [\ex(Y_{x_1, W_{x_1}} \mid z_0) - \ex(Y_{x_0, W_{x_1}} \mid z_0)] &= -1
    \end{align}
    Therefore, an interaction that was not visible in the $x\text{-DE-IE}_{x_0, x_1}(y \mid x) = 0$ effect, becomes visible once the conditioning on $X=x$ is replaced by the conditioning on $Z = z$. 
\end{example}
The above example illustrates an important concept of \textit{power}, previously discussed in \citep{plevcko2024causal}. In particular, $x\text{-DE-IE}_{x_0, x_1}(y \mid x)$ measures the direct-indirect effect interaction across all units compatible with $X(u) = x$. It can be thus be expanded as:
\begin{align}
    x\text{-DE-IE}_{x_0, x_1}(y \mid x) &= \ex \Big( [Y_{x_1, W_{x_0}}(u) - Y_{x_0, W_{x_0}}(u)] - [Y_{x_1, W_{x_1}}(u) - Y_{x_0, W_{x_1}}(u)] \mid X = x\Big), \\
    &= \sum_z \ex \Big( [Y_{x_1, W_{x_0}}(u) - Y_{x_0, W_{x_0}}(u)] - [Y_{x_1, W_{x_1}}(u) - Y_{x_0, W_{x_1}}(u)] \mid X = x, Z =z\Big) \\ &\quad \times \pr(Z = z \mid X = x) \nonumber.
\end{align}
In the causal diagram from Fig.~\ref{fig:sfm}, we have that $Y_{x, W_{x'}} \ci X \mid Z=z$ for any choice of $x, x'$, known as conditional ignorability, which is a consequence of the fact that there are no back-door paths between $X, Y$. In the SCM $\mathcal{M}$ in Eqs.~\ref{eq:scm-de-ie-power-z}-\ref{eq:scm-de-ie-power-y} of Ex.~\ref{ex:de-ie-granularity}, we also have $X \ci Z$, and hence we can write:
\begin{align}
    \hspace{-3pt} x\text{-DE-IE}_{x_0, x_1}(y \mid x) &= \sum_z \ex \Big( [Y_{x_1, W_{x_0}}(u) - Y_{x_0, W_{x_0}}(u)] - [Y_{x_1, W_{x_1}}(u) - Y_{x_0, W_{x_1}}(u)] \mid Z =z\Big) \\ &\quad\qquad
    \times \pr(Z = z \mid X = x) \nonumber\\
    &= \sum_z z\text{-DE-IE}_{x_0, x_1}(y \mid z) \pr(Z = z),
\end{align}
where $z\text{-DE-IE}_{x_0, x_1}(y \mid z) = \ex \Big( [Y_{x_1, W_{x_0}}(u) - Y_{x_0, W_{x_0}}(u)] - [Y_{x_1, W_{x_1}}(u) - Y_{x_0, W_{x_1}}(u)] \mid Z =z\Big)$. 
Therefore, $x\text{-DE-IE}_{x_0, x_1}(y \mid x)$ is a mixture of the $z$-specific direct-indirect interaction effects, which happen to cancel out in Ex.~\ref{ex:de-ie-granularity}. However, when performing interaction testing, one can also directly compute the more granular, $z$-specific effects $z\text{-DE-IE}_{x_0, x_1}(y \mid z)$ instead of the $x$-specific $x\text{-DE-IE}_{x_0, x_1}(y \mid x)$. In the language of interaction testing, we can say that the test
\begin{align}
    H^{z\text{-DE-IE}}_0: z\text{-DE-IE}_{x_0, x_1}(y \mid z) = 0,
\end{align}
is a valid interaction test for the DE-IE interaction for any fixed choice of $Z = z$, and may be superior to testing $x\text{-DE-IE}_{x_0, x_1}(y \mid x)$ against $0$. 

\paragraph{Further Population Granularity.} At the conceptual level, it is possible to consider even more granular DE-IE interactions, for instance by conditioning on $v = (x, z, w)$, or even $v = (x, z, w, y)$, by using measures of the form:
{\small
\begin{align} \label{eq:xzw-de-ie}
    v\text{-DE-IE}_{x_0, x_1}(y \mid x,z,w) &= \ex \Big( [Y_{x_1, W_{x_0}}(u) - Y_{x_0, W_{x_0}}(u)] - [Y_{x_1, W_{x_1}}(u) - Y_{x_0, W_{x_1}}(u)] \mid v\Big).
\end{align}\hspace{-3pt}} These measures quantify the DE-IE interaction for the set of units compatible with $X = x, Z = z, W= w$, or $X = x, Z = z, W= w, Y=y$. This provides an even more granular quantification of the interaction effects, for an increasingly smaller population. One may even consider interactions at the unit level
\begin{align} \label{eq:u-de-ie}
    u\text{-DE-IE}_{x_0, x_1}(y \mid u) &= [Y_{x_1, W_{x_0}}(u) - Y_{x_0, W_{x_0}}(u)] - [Y_{x_1, W_{x_1}}(u) - Y_{x_0, W_{x_1}}(u)],
\end{align}
which quantifies the effect for a single unit $u$, and can thus be seen as the most granular possible quantification of the DE-IE interaction. However, we remark that the measures in Eqs.~\ref{eq:xzw-de-ie}-\ref{eq:u-de-ie} are not identifiable from observational data even in the causal diagram in Fig.~\ref{fig:sfm}, since they would require the evaluation of the joint distribution of counterfactual outcomes, which cannot be done without much stronger (and often untestable) assumptions. 
\section{Experiments} \label{sec:experiments}
In this section, we perform an empirical evaluation of the proposed methods.
The first part of the experiments focuses on synthetic data with a known ground truth, while the second part investigates interactions in real-world datasets. 
For performing interaction testing, we use the known identification expressions for the required potential outcomes \citep{pearl:2k, plevcko2024causal}, which are valid for the causal diagram in Fig.~\ref{fig:sfm} (for brevity, we skip a detailed discussion on effect identification). We then estimate the associated measures from observational data, by deriving the efficient influence functions \citep{van2000asymptotic} for each estimator, and performing one-step bias correction to derive the estimators \citep{chernozhukov2018double, kennedy2022semiparametric}. In particular, the learners we use are based on gradient boosting \citep{chen2016xgboost}, and we perform 10-fold cross-fitting to ensure asymptotic normality of the estimators. 
All of the code used for the experiments can be found in the associated \href{https://github.com/dplecko/ia-testing}{Github repository}. 

\subsection{Synthetic Experiments}
We begin by analyzing synthetic data with a known ground truth. We consider five SCMs, labeled $\mathcal{M}_1$ to $\mathcal{M}_5$, with details given in Appendix~\ref{appendix:exp-details}. 
\begin{table}[t]
    \renewcommand{\arraystretch}{1.2}
    \centering
    \begin{tabular}{lcccccc}
\hline
Dataset & TE $\otimes$ SE & DE $\otimes$ IE & DE $\otimes$ SE & IE $\otimes$ SE & DE $\otimes$ IE $\otimes$ SE \\
\hline
$\mathcal{M}_1$ & \ding{52} & \ding{52} & \ding{52} & \ding{56} & \ding{56} \\\hline
$\mathcal{M}_2$ & \ding{52} & \ding{56} & \ding{56} & \ding{52} & \ding{56} \\\hline
$\mathcal{M}_3$ & \ding{56} & \ding{56} & \ding{56} & \ding{56} & \ding{56} \\\hline
$\mathcal{M}_4$ & \ding{52} & \ding{56} & \ding{52} & \ding{56} & \ding{56} \\\hline
$\mathcal{M}_5$ & \ding{52} & \ding{52} & \ding{52} & \ding{52} & \ding{52} \\
\hline
\end{tabular}
    \caption{Interaction summary for synthetic SCMs $\mathcal{M}_1$ to $\mathcal{M}_5$.}
    \label{tab:synthetic-ias}
\end{table}
In Tab.~\ref{tab:synthetic-ias}, we provide a summary of which interactions are active in the SCMs $\mathcal{M}_1$ to $\mathcal{M}_5$. Then, over 100 repetitions $i \in \{1, \dots, 100\}$, and over a range of sample sizes $n \in \{500, 750, 1500, 3000, 5000, 8000\}$, we generate a random dataset $\mathcal{D}(i, n)$ of $n$ samples from each SCM. In each instance, we estimate the measures associated with TE-SE, DE-IE, DE-SE, IE-SE, and DE-IE-SE interactions using data $\mathcal{D}(i, n)$. Based on asymptotic normality, we also derive the p-values for each interaction test (recall that we are interested in whether the interaction measure $\mathcal{C} = 0$). Whenever there is no interaction, meaning that $\mathcal{C} = 0$ (on the population level) and the null-hypothesis $H_0$ for an interaction test (Def.~\ref{def:ia-test}) is true, we expect the distribution of the p-values to be approximately uniform. When $H_0$ is not true, we expect the distribution of the p-values to be shifted towards zero, which would indicate power for detecting an interaction. 

\begin{figure}
    \centering
    \includegraphics[width=\linewidth]{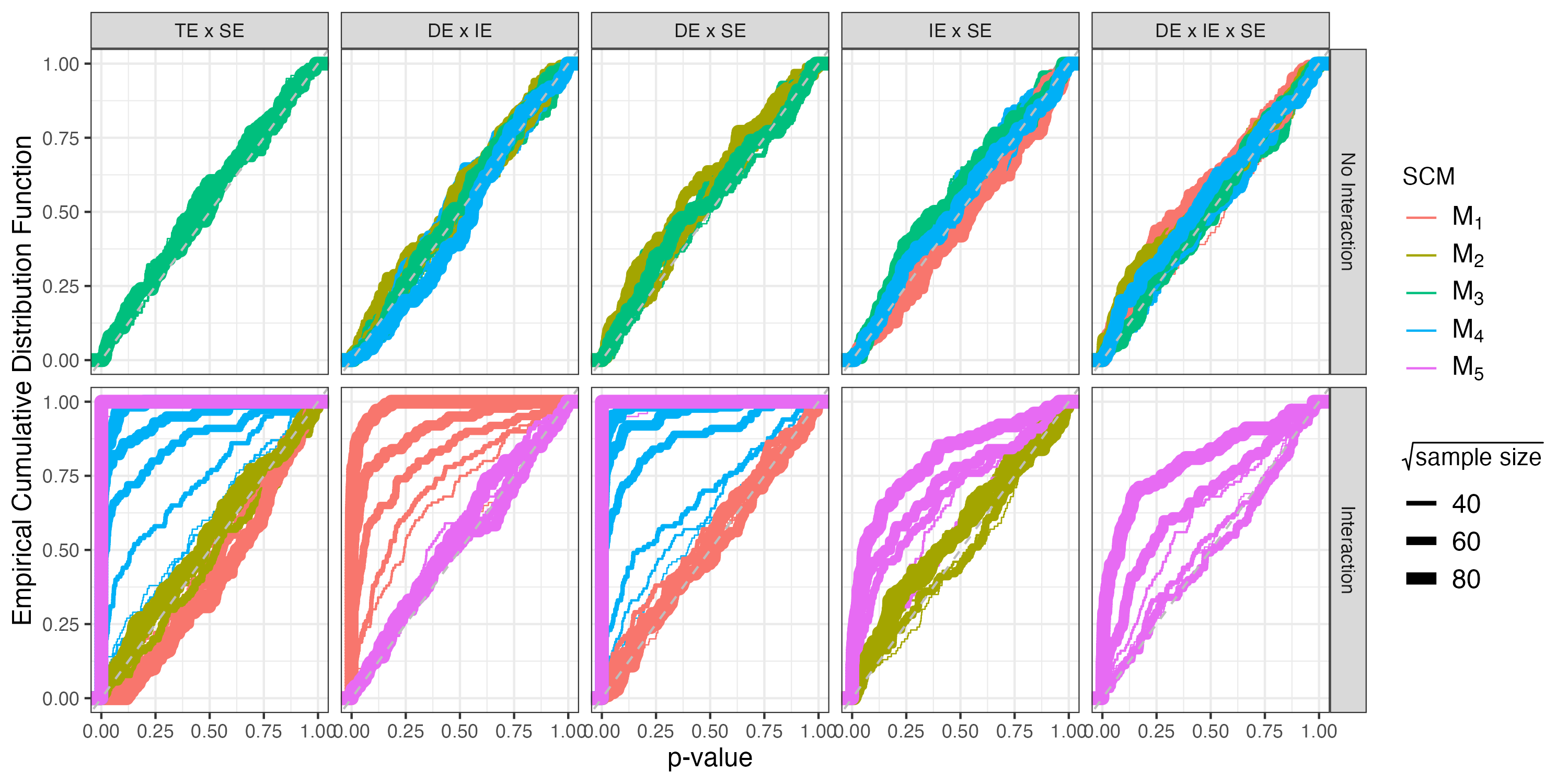}
    \caption{Distribution of p-values for interaction testing.}
    \label{fig:p-values}
\end{figure}

In Fig.~\ref{fig:p-values}, we plot the empirical cumulative distribution functions (ECDF) of the p-values obtained, stratified by the SCM $\mathcal{M}_i$, sample size $n$, and the type of interaction. In the top row of the figure, we plot the ECDFs for interactions and SCMs where the null-hypothesis (no interaction) holds true. As mentioned, under the null, we expect the distribution of the p-values to be approximately uniform, which holds true for the data at hand. 
Formally, however, a Kolmogorov-Smirnov test comparing the ECDF of the p-values with the Unif$[0,1]$ distribution rejects the null-hypothesis of the uniform distribution with $p < 0.001$.

For the interactions and SCMs where the null-hypothesis of no interaction is false, we expect our interaction tests to provide some level of power for detecting the violation of $H_0$. Such power would be reflected in the shift of the distribution of p-values towards $0$, which is observed in the second row of Fig.~\ref{fig:p-values}. As expected, with increased sample size power also increases  (corresponding to thicker lines in the plot). We also remark that not all interactions can be detected with sample sizes up to 8000, tying back to the discussion on granularity in Sec.~\ref{sec:population-axis}. In a nutshell, even though an interaction may be present, the corresponding measure for the interaction test may be close or equal to $0$, making the testing problem difficult. In Fig.~\ref{fig:alpha-testing} of Appendix~\ref{appendix:exp-details}, we illustrate the results of the hypothesis testing at the level $\alpha = 0.05$, showing that the tests attain the correct significance level and increasing power with increased sample size.

\subsection{Real-World Data}
The goal of the second part of the evaluation is to understand how often we detect different types of interactions. The experimental part is intended to answer the question of ``how often do causal pathways interact in real-world datasets?''. For answering this question, we use a selection of 10 datasets including data in criminal justice, economics, marketing, banking, medical treatments, epidemiology, and public health. 
For each dataset, we first construct the graphical model shown in Fig.~\ref{fig:sfm}. This is done by selecting the $X, Z, W,$ and $Y$ variable sets. 
Again, we consider the five types of interaction as before, namely the interactions of spurious and total effects, and direct and indirect effects, appearing in the decompositions in Thm.~\ref{thm:1st-tv-decomp} and Thm.~\ref{thm:1st-tvr-decomp}. 
Recall that at the structural level these interactions were determined based to the existence of different terms in the structural mechanisms $f_w, f_y$ and the existence of different paths between $X, Y$ (Def.~\ref{def:str-ia}). 
Furthermore, we also consider the more granular interactions, including spurious-direct, spurious-indirect, and direct-indirect-spurious (Def.~\ref{def:str-ia-granular}).
For each interaction, we use a corresponding interaction test for testing the existence of the interaction at: (i) difference scale; (ii) log-risk difference scale; (iii) odds ratio scale (only for binary outcomes $Y$).  
Below we provide a short description of the datasets for experiments, including the description of how variables $X, Z, W$, and $Y$ were chosen, while the \href{https://github.com/dplecko/ia-testing/blob/main/r/load-data.R}{source code} gives the full list of variables used for each dataset:
\begin{enumerate}[label=(\roman*)]
    \item \textbf{COMPAS:} \citep{larson2016how} The COMPAS dataset contains information used for predicting recidivism risk among individuals seeking parole. Variables include demographic information (Z), recidivism outcome (Y), prior criminal history (W) and race (X). The focus is on understanding the association of race and the recidivism risk.
    
    \item \textbf{Census 2018:} \citep{cevid2022distributional, plevcko2024causal} The dataset contains detailed demographic information from the 2018 US Census. Variables include sex (X), income level in dollars per year (Y), employment status and educational background (W), and other demographic variables (Z). The focus is on understanding the association of sex and income level.

    \item \textbf{UCI Credit:} \citep{uci2016credit} The UCI Credit dataset contains information about default of credit card payments of bank customers in Taiwan. Variables include applicant sex (X), default on payment (Y), age (Z), and education, marital status, and recent financial behavior (W). The focus is on understanding the association of customer sex and default on payment.
    
    \item \textbf{MIMIC-IV:} \citep{johnson2023mimic} The MIMIC-IV dataset contains electronic health records from a large tertiary hospital center in Boston, Massachusetts. We are interested in patients admitted to the intensive care unit (ICU). Variables include patients' race (X), mortality outcome (Y), comorbidities and demographic information (Z), and illness severity and treatment information (W). The focus is on understanding the association of race and mortality in intensive care unit outcomes.
    
    \item \textbf{HELOC:} \citep{heloc2016fico} The Home Equity Line of Credit (HELOC) dataset is provided by the Fair Isaac Corporation (FICO) and includes credit applications made by prospective homeowners. Variables include proportion of transaction delinquencies (X), credit risk performance (Y), financial background and debt history (Z), and recent credit utilization (W). The focus is on understanding the association of high number of transactions delinquencies and the estimate of the credit risk performance.
    
    \item \textbf{UCI Adult:} \citep{uci1996adult} This dataset includes census data from the United States for predicting whether the individual's income exceeds \$50K/yr. Variables include education level (X), income level (Y), demographic factors (Z), and family/employment characteristics (W). The focus is on understanding the association of high levels of education and income level.
    
    \item \textbf{UCI Wine Quality:} \citep{cortez2009modeling} This dataset consists of physicochemical tests of wine samples and their quality ratings. Variables include alcohol content (X), wine quality (Y), chemical composition (Z), and physical/fermentation properties (W). The focus is on understanding the association of high alcohol content and wine quality.
    
    \item \textbf{UCI Bank Marketing:} \citep{uci2012bank} This dataset includes marketing campaign data from a Portuguese banking institution. Variables include housing loan status (X), agreement for a term deposit subscription (Y), demographic information (Z), and campaign details (W). The focus is on understanding the association of housing loan status and term deposit subscription.
    
    \item \textbf{UCI Estimation of Obesity:} \citep{palechor2019dataset} This dataset contains information on Colombian individuals' eating habits and physical attributes to determine obesity levels. Variables include family history of overweight (X), body mass index (BMI, labeled Y), demographic factors (Z), and lifestyle habits (W). The focus is on understanding the association of family history of overweight and BMI.
    
    \item \textbf{BRFSS Diabetes Health Indicators:} This dataset was collected by Behavioral Risk Factor Surveillance System (BRFFSS), and contains health-related survey data to assess diabetes risk factors. Variables include physical activity (X), diabetes status (Y), demographic and lifestyle factors (Z), and health conditions (W). The focus is on understanding the association of physical activity and diabetes status.
\end{enumerate}
\begin{table}[t]
    \renewcommand{\arraystretch}{1.2}
    \centering
    \begin{tabular}{|c|c|c|c|c|c|c|}
        \hline
        \backslashbox{Dataset}{Interaction} & samples $n$ & TE $\otimes$ SE & DE $\otimes$ IE & DE $\otimes$ SE & IE $\otimes$ SE & DE $\otimes$ IE $\otimes$ SE \\ \hline
COMPAS & 7214 &  &  &  &  &   \\ \hline 
Census 2018 & 20000 & \textbullet & \textbullet &  &  &   \\ \hline 
UCI Credit & 30000 &  &  &  &  &   \\ \hline 
MIMIC-IV & 38844 & \textbullet &  &  & \textbullet &   \\ \hline 
HELOC & 10459 & \textbullet & \textbullet & \textbullet &  &   \\ \hline 
UCI Adult & 48842 &  & \textbullet &  &  &   \\ \hline                                                        
UCI Wine Quality & 6497 & \textbullet &  &  &  &   \\ \hline 
UCI Bank Marketing & 45211 &  &  &  &  &   \\ \hline 
UCI Obesity & 2111 &  & \textbullet &  &  &   \\ \hline 
BRFSS Diabetes & 20000 &  &  &  &  &   \\ \hline 
    \end{tabular}
    \caption{Interaction testing on 10 real-world datasets.}
    \label{tab:real-data}
\end{table}
The summary of the experimental results is shown in Tab.~\ref{tab:real-data}. Overall, we found that 10/50 of the investigated interactions seemed to be significant at the 5\% level (when applying the Benjamini-Hochberg procedure \citep{benjamini1995controlling}, three hypotheses were deemed significant when controlling the false discovery rate at the 5\% level). 
We remark that the interaction tests performed here are not the most powerful possible (i.e., more granular tests could be used, as discussed in Sec.~\ref{sec:population-axis}). In practice, when interactions are significant, the analyst can include them explicitly in the TV decomposition. If interactions are not significantly different from $0$, then a more parsimonious version of the TV decomposition may be used (Alg.~\ref{algo:ia-testing}).
\section{Conclusion}
In this paper, we argued that a new concept of \textit{variation analysis} should be adopted, to reflect recent methodological advances in which confounded/spurious effects are considered when analyzing co-variations between a treatment $X$ and an outcome $Y$. Traditionally, in mediation analysis, only the causal variations (direct, indirect) originating from $X$ and entering $Y$ are considered, and the quantity that is under study is often the total effect (TE, which encompasses direct and indirect effects). In variation analysis, however, the focus is on the total variation measure (TV, which encompasses all co-variations between $X$, $Y$), thereby representing a (strictly) more general approach than mediation analysis.
We further introduced the concept of structural interactions (Defs.~\ref{def:str-ia}, \ref{def:str-ia-granular}), and discussed how different causal pathways (direct, indirect, and spurious) between treatment $X$ and outcome $Y$ can interact, and how such interactions can be quantified (Def.~\ref{def:xspecific}). 
We then proved decomposition results for the TV measure that include and quantify all the different interactions among causal paths (Thms.~\ref{thm:1st-tv-decomp}, \ref{thm:2nd-tv-decomp}), and require only $x_0 \to x_1$ transitions in all the effects, allowing for an easier interpretation of the decomposition. 
We then defined the concept of interaction testing, a procedure aimed at testing the existence of interactions between different causal paths (Def.~\ref{def:ia-test}). Subsequently, we demonstrated that whenever a hypothesis is not rejected, it implies that the decomposition of the TV measure can be made more parsimonious, by omitting the corresponding interaction term (Alg.~\ref{algo:ia-testing}). Once again, this provides a practical approach that allows users to more easily interpret the decomposition of the TV measure.
Further, we extended the above results to include the log-risk and log-odds scales (Appendix.~\ref{appendix:lr-lo-scales}).
Finally, in Sec.~\ref{sec:experiments}, we performed an empirical analysis of our method on synthetic data with a known ground truth, to illustrate its validity. We also performed an extensive analysis of 10 well-known and commonly used datasets -- in an attempt to discover how often there are significant interactions among causal pathways. Over 10 datasets and 5 types of interactions, we found that 10 out of 50 possible interactions were significant at the 5\% significance level (without multiple testing adjustments). Therefore, it may often be possible to use more parsimonious decompositions of the TV measure. Whenever this is not possible (the no interaction hypothesis is rejected), the analyst may include the interaction effect explicitly in the decomposition, and further investigate the significant interaction (see Sec.~\ref{sec:population-axis}).

\newpage
\bibliography{refs}

\appendix
\section{Theorem Proofs}
This appendix contains the proof for the main theorems and results appearing in the main text.

\begin{proof}[Thm.~\ref{thm:1st-tv-decomp}]
    We first prove the statement in Eq.~\ref{eq:tv-decomp-te-se}. Note that we have:
    \begin{align}
        x\text{-TE-SE}_{x_0, x_1}(y) &= [\ex(Y_{x_1} \mid x_1) - \ex(Y_{x_0} \mid x_1)] - [\ex(Y_{x_1} \mid x_0) - \ex(Y_{x_0} \mid x_0)] \\
        &= \underbrace{\ex(Y_{x_1} \mid x_1) - \ex(Y_{x_0} \mid x_0)}_{\text{TV}_{x_0, x_1}(y)} + \underbrace{\ex(Y_{x_0} \mid x_0) - \ex(Y_{x_0} \mid x_1)}_{-x\text{-SE}_{x_0, x_1}(y)} - x\text{-TE}_{x_0, x_1}(y \mid x_0) \\
        &= \text{TV}_{x_0, x_1}(y) - x\text{-SE}_{x_0, x_1}(y) - x\text{-TE}_{x_0, x_1}(y \mid x_0),
    \end{align}
    which yields Eq.~\ref{eq:tv-decomp-te-se} after rearrangement. To prove Eq.~\ref{eq:tv-decomp-de-ie-se}, we show that
    \begin{align}
        x\text{-TE}_{x_0, x_1}(y \mid x_0) &= \ex(Y_{x_1} \mid x_0) - \ex(Y_{x_0} \mid x_0) \\
        &= \ex(Y_{x_1} \mid x_0) - \ex(Y_{x_1, W_{x_0}} \mid x_0) + \underbrace{\ex(Y_{x_1, W_{x_0}} \mid x_0) - \ex(Y_{x_0} \mid x_0)}_{x\text{-DE}_{x_0, x_1}(y \mid x_0)} \\
        &= \underbrace{\ex(Y_{x_1} \mid x_0) - \ex(Y_{x_1, W_{x_0}} \mid x_0) - [\ex(Y_{x_0, W_{x_1}} \mid x_0) - \ex(Y_{x_0, W_{x_0}} \mid x_0)]}_{x\text{-DE-IE}_{x_0, x_1}(y \mid x_0)} \\
        &\quad + \underbrace{\ex(Y_{x_0, W_{x_1}} \mid x_0) - \ex(Y_{x_0, W_{x_0}} \mid x_0)}_{x\text{-IE}_{x_0, x_1}(y \mid x_0)} + x\text{-DE}_{x_0, x_1}(y \mid x_0) \\
        &= x\text{-DE}_{x_0, x_1}(y \mid x_0) + x\text{-IE}_{x_0, x_1}(y \mid x_0) + x\text{-DE-IE}_{x_0, x_1}(y \mid x_0).
    \end{align}
    Using the above in Eq.~\ref{eq:tv-decomp-te-se} yields Eq.~\ref{eq:tv-decomp-de-ie-se}, completing the theorem's proof.
\end{proof}

\begin{proof}[Thm.~\ref{thm:2nd-tv-decomp}]
    Following the definitions of interactions in Eqs.~\ref{eq:x-de-se}-\ref{eq:x-de-ie-se}, we have that
    \begin{align} 
        &x\text{-DE-SE}_{x_0, x_1}(y) + x\text{-IE-SE}_{x_0, x_1}(y) + x\text{-DE-IE-SE}_{x_0, x_1}(y) \\  
        &= [\underbrace{\ex(Y_{x_1, W_{x_0}} \mid x_1) - \ex(Y_{x_0} \mid x_1)}_{T_1}] - [\underbrace{\ex(Y_{x_1, W_{x_0}} \mid x_0) - \ex(Y_{x_0} \mid x_0)}_{T_2}] \\ 
        &\quad + [\underbrace{\ex(Y_{x_0, W_{x_1}} \mid x_1)}_{T_3} - \ex(Y_{x_0} \mid x_1)] - [\underbrace{\ex(Y_{x_0, W_{x_1}} \mid x_0)}_{T_4} - \ex(Y_{x_0} \mid x_0)] \\
        &\quad + [\ex(Y_{x_1} \mid x_1) - \underbrace{\ex(Y_{x_0, W_{x_1}} \mid x_1)}_{T_3}] - [\underbrace{\ex(Y_{x_1, W_{x_0}} \mid x_1) - \ex(Y_{x_0} \mid x_1)}_{T_1}] \\ 
        &\quad 
        - \big[ [\ex(Y_{x_1} \mid x_0) - \underbrace{\ex(Y_{x_0, W_{x_1}} \mid x_0)}_{T_4}] - [\underbrace{\ex(Y_{x_1, W_{x_0}} \mid x_0) - \ex(Y_{x_0} \mid x_0)}_{T_2}] \big] \\
        &= -\ex(Y_{x_0} \mid x_1) + \ex(Y_{x_0} \mid x_0) + \ex(Y_{x_1} \mid x_1) - \ex(Y_{x_1} \mid x_0) \\
        &= [\ex(Y_{x_1} \mid x_1) - \ex(Y_{x_0} \mid x_1)] - [\ex(Y_{x_1} \mid x_0) - \ex(Y_{x_0} \mid x_0)] \\
        &= x\text{-TE-SE}_{x_0, x_1}(y),
    \end{align}
    where in the second step, terms $T_1, \dots, T_4$ cancel out due to opposite signs. Using the above equality in Eq.~\ref{eq:tv-decomp-de-ie-se} yields the decomposition in Eqs.~\ref{eq:tv-first-order}-\ref{eq:tv-third-order}.
\end{proof}
Proving Thm.~\ref{thm:1st-tvr-decomp} of Appendix~\ref{appendix:lr-lo-scales} follows exactly the same steps as the proof of Thm.~\ref{thm:1st-tv-decomp}. The only difference is that one must (i) replace the potential outcome $Y_C$ with $P_C$; (ii) log-transform the outcome to $\log P_C$. All other steps of the proofs remain the same.

\section{Assessing Structural Interactions} \label{appendix:str-decision-trees}
\begin{figure}[t]
    \centering
    \begin{tikzpicture}[every node/.style={rectangle, draw, minimum width=1.2cm}, 
                    grow=south, sibling distance=5cm, level distance=2cm,
                    edge from parent/.style={draw, -Latex, thick}]

    \tikzset{labnode/.style={midway, above, yshift=0.2cm, draw=none}}
    \node {X $\leftarrow$ Z}
        child { node {\ding{56}}
            edge from parent node[labnode] {no}
        }
        child { node {W $\leftarrow$ X}
            child { node {\ding{56}}
                edge from parent node[labnode] {no}
            }
            child { node {$f_w(X, Z, U_w)$}
                child { node {$f_y(Z, W, U_y)$}
                    [sibling distance=3cm]
                    child { node {$f_y$ non-linear in $W$}
                        child {
                            node {\ding{56}}
                            edge from parent node[labnode] {no}
                        }
                        child {
                            node {\ding{52}}
                            edge from parent node[labnode] {yes}
                        }
                        edge from parent node[labnode] {no}
                    }
                    child { node {\ding{52}} 
                        edge from parent node[labnode] {yes}
                    }
                    edge from parent node[labnode] {no}
                }
                child { node {$W \rightarrow Y$}
                    [sibling distance=3cm]
                    child { node {\ding{56}} 
                        edge from parent node[labnode] {no}
                    }
                    child { node {\ding{52}} 
                        edge from parent node[labnode] {yes}
                    }
                    edge from parent node[labnode] {yes}
                }
                edge from parent node[labnode] {yes}
            }
            edge from parent node[labnode] {yes}
        };
\end{tikzpicture}
    \caption{Assessing structural interaction of indirect and spurious effects (Str-IE-SE).}
    \label{fig:str-ie-se-flow}
\end{figure}
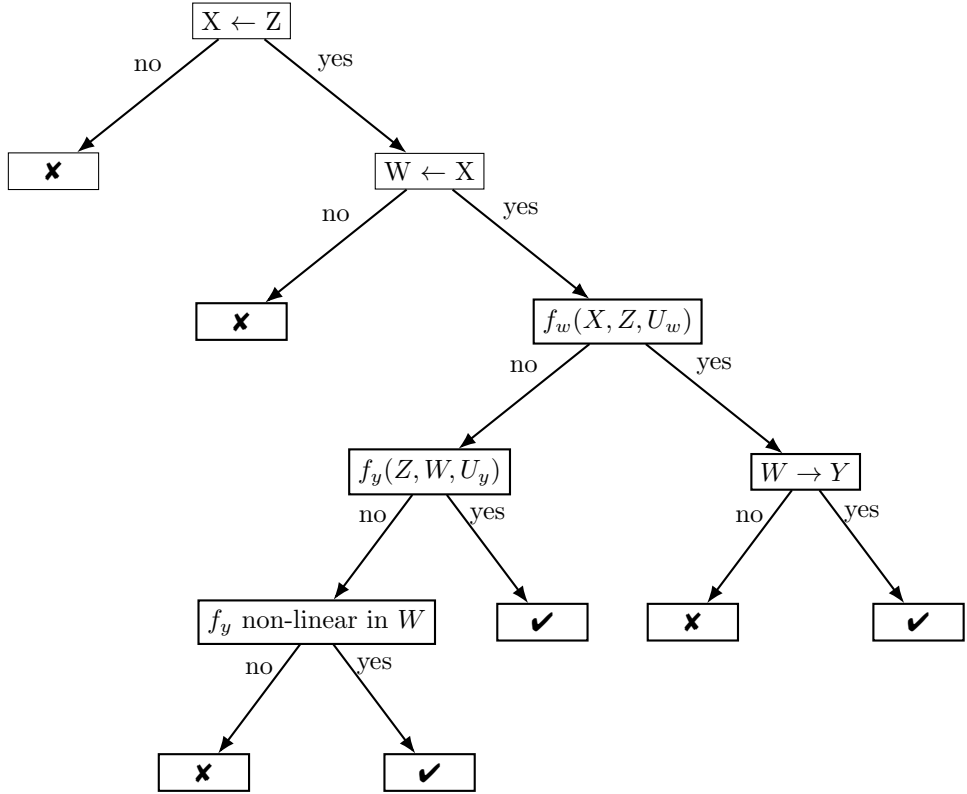

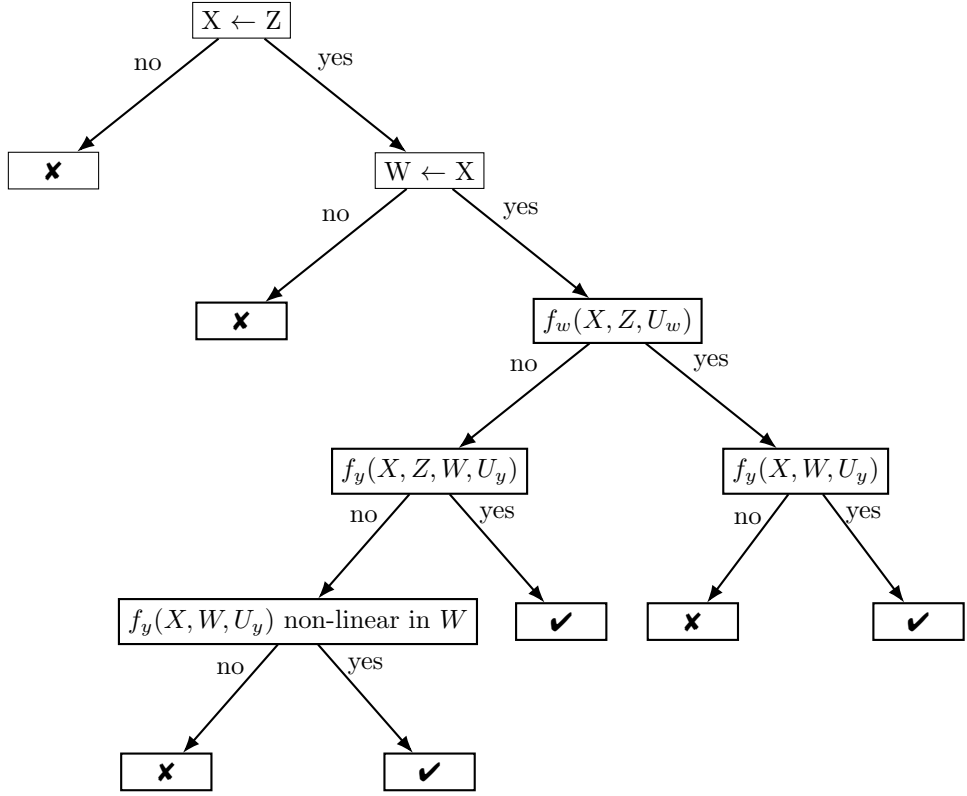
\begin{figure}[t]
    \centering
    \begin{tikzpicture}[every node/.style={rectangle, draw, minimum width=1.2cm}, 
                    grow=south, sibling distance=5cm, level distance=2cm,
                    edge from parent/.style={draw, -Latex, thick}]

    \tikzset{labnode/.style={midway, above, yshift=0.2cm, draw=none}}
    \node {X $\leftarrow$ Z}
        child { node {\ding{56}}
            edge from parent node[labnode] {no}
        }
        child { node {W $\leftarrow$ X}
            child { node {\ding{56}}
                edge from parent node[labnode] {no}
            }
            child { node {$f_w(X, Z, U_w)$}
                child { node {$f_y(X, Z, W, U_y)$}
                    [sibling distance=3.5cm]
                    child { node {$f_y(X, W, U_y)$ non-linear in $W$}
                        child {
                            node {\ding{56}}
                            edge from parent node[labnode] {no}
                        }
                        child {
                            node {\ding{52}}
                            edge from parent node[labnode] {yes}
                        }
                        edge from parent node[labnode] {no}
                    }
                    child { node {\ding{52}} 
                        edge from parent node[labnode] {yes}
                    }
                    edge from parent node[labnode] {no}
                }
                child { node {$f_y(X, W, U_y)$}
                    [sibling distance=3cm]
                    child { node {\ding{56}} 
                        edge from parent node[labnode] {no}
                    }
                    child { node {\ding{52}} 
                        edge from parent node[labnode] {yes}
                    }
                    edge from parent node[labnode] {yes}
                }
                edge from parent node[labnode] {yes}
            }
            edge from parent node[labnode] {yes}
        };
\end{tikzpicture}
    \caption{Assessing structural interaction of direct, indirect, and spurious effects (Str-DE-IE-SE).}
    \label{fig:str-de-ie-se-flow}
\end{figure}

In this appendix, we provide an expanded discussion of the structural interaction notions from Defs.~\ref{def:str-ia} and \ref{def:str-ia-granular}. In fact, while the definitions of structural interactions may seem involved, containing multiple conditions, there is a systematic way of deriving these definitions, which is now discussed. In particular, Figs.~\ref{fig:str-de-se-flow}, \ref{fig:str-ie-se-flow}, and \ref{fig:str-de-ie-se-flow} provide specific decision trees that allow one to assess the existence of DE-SE, IE-SE, and DE-IE-SE interactions, respectively.

\begin{figure}[t]
    \centering
    \begin{tikzpicture}[every node/.style={rectangle, draw, minimum width=1.2cm}, 
                    grow=south, sibling distance=5cm, level distance=2cm,
                    edge from parent/.style={draw, -Latex, thick}]

    \tikzset{labnode/.style={midway, above, yshift=0.2cm, draw=none}}
    \node {X $\leftarrow$ Z}
        child { node {\ding{56}}
            edge from parent node[labnode] {no}
        }
        child { node {W $\leftarrow$ Z}
            child { node {$f_y(X, Z, U_y)$}
                [sibling distance=3cm]
                child { node {\ding{56}} edge from parent node[labnode] {no}}
                child { node {\ding{52}} edge from parent node[labnode] {yes}}
                edge from parent node[labnode] {no}
            }
            child { node {$f_y(X, W, U_y)$ or $f_y(X, Z, U_y)$}
                [sibling distance=3cm]
                child { node {\ding{56}} edge from parent node[labnode] {no}}
                child { node {\ding{52}} edge from parent node[labnode] {yes}}
                edge from parent node[labnode] {yes}
            }
            edge from parent node[labnode] {yes}
        };
\end{tikzpicture}
    \caption{Assessing structural interaction of direct and spurious effects (Str-DE-SE).}
    \label{fig:str-de-se-flow}
\end{figure}
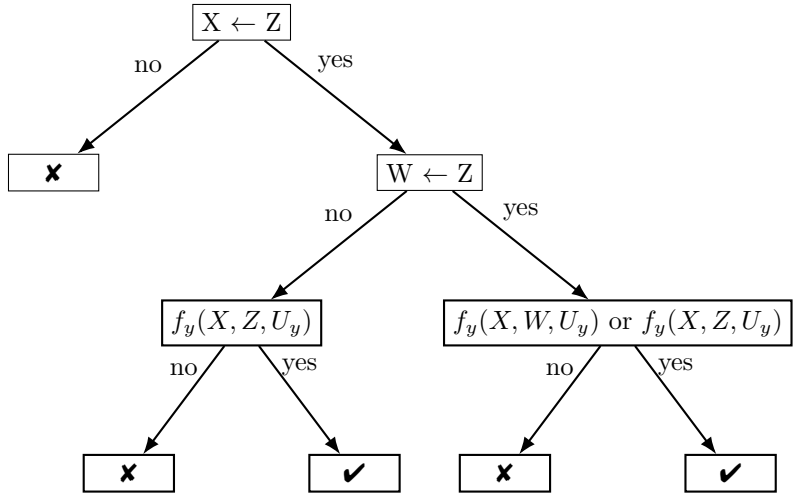

To provide the explanation of how to interpret these decision trees, we discuss the one in Fig.~\ref{fig:str-ie-se-flow} in detail, and the other two can be understood along very similar lines. The decision tree in Fig.~\ref{fig:str-ie-se-flow} provides the reasoning for assessing the interaction of spurious and indirect pathways on the structural level (IE-SE interaction). In the first step, we ask whether the structural mechanism of $X$ takes $Z$ as an input (written $X \leftarrow Z$ for short). If not, we can conclude that there is no spurious-indirect interaction, since if $X$ does not take $Z$ as input, there are no open back-door paths between $X$ and $Y$. If yes, we move onto the next node, and verify whether $f_w$ takes $X$ as an input. If not, there can be no interaction between spurious and indirect effects, since there is no indirect path from $X$ to $Y$. Otherwise, we check whether there exists a term in $f_w$ that requires both $X, Z$ as inputs (this indicates whether $X$ and $Z$ interaction in $W$). 

If not, we further check if a term in $f_y$ exists that requires $W, Z$ as inputs. If not, we conclude that there is no interaction of the effects, since $X, Z$ do not interact at $W$, nor $Z, W$ interact at $Y$. 

If $X$ and $Z$ do interact at $W$, however, we next check whether $W$ is an input argument of $f_y$. If yes, this gives an interaction of pathways, since $X \leftarrow Z \rightarrow W$ and $X \rightarrow W$ interact, and this interaction is transmitted to $Y$ along the $W \to Y$ arrow. Finally, if $W$ is not an input of $f_y$, then the interaction of $X \leftarrow Z \rightarrow W$ and $X \rightarrow W$ cannot be transmitted to $Y$, meaning there is no interaction of spurious and indirect pathways.

\section{Structural Admissibility of Interaction Measures}
We now prove the admissibility results for the interaction measures from Defs.~\ref{def:xspecific} and \ref{def:xspecific-granular} with respect to the structural interaction criteria from Defs.~\ref{def:str-ia} and \ref{def:str-ia-granular}. \\
\begin{proof}[Props.~\ref{prop:str-admissibility} and \ref{prop:str-admissibility-granular}]
    We deal with each interaction separately, and proceed as follows. For any condition that provides a no-interaction result for a structural criterion, we prove that the corresponding interaction measure is 0. Whenever the no-interaction structural criteria are not satisfied, we provide examples that justify and ground the intuition for why the structural interaction is present.

    \paragraph{TE-SE Interaction.} Suppose that condition \ref{cond:no-f-xz} from Def.~\ref{def:str-ia} holds true. Then, for any $U = u$, we have that
    \begin{align}
        Y_{x_1}(u) - Y_{x_0}(u) &= f^{(1)}_y(x_1, u_y) + f^{(2)}_y(Z(u), u_y) - (f^{(1)}_y(x_0, u_y) + f^{(2)}_y(Z(u), u_y)) \\
        &= f^{(1)}_y(x_1, u_y) - f^{(1)}_y(x_0, u_y).
    \end{align}
    Therefore, we can write
    \begin{align}
        \ex[Y_{x_1} - Y_{x_0} \mid X = x] &= \ex[f(x_1, Z, U_y) - f(x_0, Z, U_y) \mid X = x] \\
        &= \ex[f(x_1, U_y) - f(x_0, U_y) \mid X = x] \\
        &= \ex[f(x_1, U_y) - f(x_0, U_y) \mid X = x'] \label{eq:uy-ci-x} \\
        &= \ex[Y_{x_1} - Y_{x_0} \mid X = x'],
    \end{align}
    where Eq.~\ref{eq:uy-ci-x} follows from the independence $U_y \ci X$. Based on this, we have that
    \begin{align}
        x\text{-TE-SE}_{x_0, x_1}(y) = \ex[Y_{x_1} - Y_{x_0} \mid X = x_1] - \ex[Y_{x_1} - Y_{x_0} \mid X = x_0] = 0.
    \end{align}
    For condition \ref{cond:z-to-x} of Def.~\ref{def:str-ia}, i.e., when there is no $X \gets Z$ edge, we know that there are no open back-door paths between $X, Y$, which implies the independence statement $Y_x \ci X$ using the back-door criterion \citep{pearl:2k}. Therefore, $\ex[Y_{x'} \mid X = x] = \ex[Y_{x'} \mid X = x'']$ for any $x, x', x''$, again implying that $x\text{-TE-SE}_{x_0, x_1}(y) = 0$. To exemplify an SCM in which the interaction is present in absence of conditions \ref{cond:no-f-xz} and \ref{cond:z-to-x}, we use the following:
    \begin{empheq}[left={\mathcal{M} = \empheqlbrace}]{align} 
  Z &\gets \text{Bernoulli}(0.5) \label{eq:te-se-scm-1}\\
  X &\gets \text{Bernoulli}(0.5 + 0.2Z) \\
  Y &\gets X + Z + XZ. \label{eq:te-se-scm-3}
\end{empheq}
For $\mathcal{M}$, we can compute:
\begin{align}
    \ex[Y_{x_1} - Y_{x_0} \mid X = x_1] &= \ex[Y_{x_1} - Y_{x_0} \mid Z = 1] P(Z = 1 \mid X = x_1) \\&\quad + \ex[Y_{x_1} - Y_{x_0} \mid Z = 0] P(Z = 0 \mid X = x_1) \\
    &= 2 \cdot \frac{6}{11} + 1 \cdot \frac{5}{11} = \frac{17}{11} \\
    \ex[Y_{x_1} - Y_{x_0} \mid X = x_0] &= \ex[Y_{x_1} - Y_{x_0} \mid Z = 1] P(Z = 1 \mid X = x_0) \\&\quad + \ex[Y_{x_1} - Y_{x_0} \mid Z = 0] P(Z = 0 \mid X = x_0) \\
    &= 2 \cdot \frac{4}{9} + 1 \cdot \frac{5}{9} = \frac{13}{9} \\
    \implies & \quad x\text{-TE-SE}_{x_0, x_1}(y) = \frac{17}{11} - \frac{13}{9} = \frac{10}{99}.
\end{align}

\paragraph{DE-IE Interaction.} For condition \ref{cond:no-f-xw} of Def.~\ref{def:str-ia}, i.e., when there is no interaction term between $X, W$ in $f_y$, we can write
\begin{align}
    Y_{x, W_{x'}}(u) &= f_y(x, W_{x'}(u), Z(u), u_y) \\
                        &= f_y^{(1)}(x, Z(u), u_y) + f_y^{(2)}(W_{x'}(u), Z(u), u_y) \\
    \implies & Y_{x_1, W_{x'}}(u) - Y_{x_0, W_{x'}}(u) = f_y^{(1)}(x_1, Z(u), u_y) - f_y^{(1)}(x_0, Z(u), u_y) \\
    & \qquad\qquad\qquad\qquad\qquad\; = Y_{x_1, W_{x}}(u) - Y_{x_0, W_{x}}(u) \\
    \implies & x\text{-DE-IE}_{x_0, x_1}(y \mid x) = 0.
\end{align}
For condition \ref{cond:x-to-w}, whenever $X$ is not an input of $f_w$, note that for any $U = u$, we have
\begin{align}
    &W_{x_0}(u) = W_{x_1}(u) \\
    &\implies Y_{x, W_{x_0}}(u) = Y_{x, W_{x_1}}(u) \;\forall x\\
    &\implies \ex[Y_{x, W_{x_1}} - Y_{x, W_{x_0}} \mid X = x'] = 0 \; \forall x, x' \\
    &\implies x\text{-DE-IE}_{x_0, x_1}(y \mid x) = 0 \;\forall x.
\end{align}
To gain a better understanding of the DE-IE interaction when conditions  \ref{cond:no-f-xw} and \ref{cond:x-to-w} are not satisfied, consider the SCM $\mathcal{M}$ given as follows:
\begin{empheq}[left={\mathcal{M} = \empheqlbrace}]{align} 
  X &\gets \text{Bernoulli}(0.5) \\
  W &\gets \text{Bernoulli}(0.5 + 0.1X)\\
  Y &\gets X + W + XW.
\end{empheq}
In this SCM, the pathways interact (by definition of Str-DE-IE), and we can also compute that $x\text{-DE-IE}_{x_0, x_1}(y \mid x) \neq 0$. \\

\noindent For the more granular interactions (DE-SE, IE-SE, DE-IE-SE), we use the tree-based representations in Figs.~\ref{fig:str-de-se-flow}-\ref{fig:str-de-ie-se-flow} to provide structure to the proofs.
\paragraph{DE-SE Interaction.} We start with the DE-SE interaction and Fig.~\ref{fig:str-de-se-flow}. First, suppose that there is no $X \gets Z$ edge. This implies there is no back-door path between $X, Y$, and consequently that $Y_{x', W_x} \ci X \;\forall x, x'.$ Thus, we have that
\begin{align}
    \ex[Y_{x', W_{x}} \mid X = x_1] = \ex[Y_{x', W_{x}} \mid X = x_0],
\end{align}
which implies that
\begin{align}
    &\ex[Y_{x_1, W_{x_0}} \mid X = x_1] = \ex[Y_{x_1, W_{x_0}} \mid X = x_0] \\
    &\ex[Y_{x_0, W_{x_0}} \mid X = x_1] = \ex[Y_{x_0, W_{x_0}} \mid X = x_0] \\
    &\qquad\qquad \implies x\text{-DE-SE}_{x_0, x_1}(y) = 0.
\end{align}
For the remainder, suppose that $X \to Z$ edge exists. We split the proof into two subparts now, depending on the existence of the $W \gets Z$ edge. If there is no $W \gets Z$ edge, we check if a term $f_y(X, Z, U_y)$ exists. If not, we can write
\begin{align}
    &Y_{x', W_x}(u) = f_y^{(1)}(x', W_x(u), u_y) + f_y^{(2)}(W_{x}(u), Z(u), u_y) \\
    &\implies Y_{x', W_x}(u) - Y_{x, W_x}(u) = f_y^{(1)}(x', W_x(u), u_y) - f_y^{(1)}(x, W_x(u), u_y).
\end{align}
Note that
\begin{align}
    \ex[Y_{x', W_x} - Y_{x, W_x} \mid x_0] &= \ex[f_y^{(1)}(x', W_x(u), u_y) - f_y^{(1)}(x, W_x(u), u_y) \mid x_0] \label{eq:de-x0} \\
    &= \ex[f_y^{(1)}(x', W_x(u), u_y) - f_y^{(1)}(x, W_x(u), u_y) \mid x_1] \label{eq:de-x1} \\
    &= \ex[Y_{x', W_x} - Y_{x, W_x} \mid x_1]
\end{align}
where Eq.~\ref{eq:de-x1} follows from $U_y, W_x \ci X$ (note that $W_x \ci X$ because there is no edge $Z \to W$ and hence no back-door path between $X, W$). Eqs.~\ref{eq:de-x0}-\ref{eq:de-x1} imply that $x\text{-DE-SE}_{x_0, x_1}(y) = 0$. 
To illustrate the setting where the interaction is present, we can again use the SCM $\mathcal{M}$ in Eqs.~\ref{eq:te-se-scm-1}-\ref{eq:te-se-scm-3}, where a term $f_y(X, Z, U_y)$ is present. For this SCM, we can also compute $x\text{-DE-SE}_{x_0, x_1}(y) = x\text{-TE-SE}_{x_0, x_1}(y) \neq 0$.

Consider now the case in which there are both $Z \to X$ and $Z \to W$ edges present. If there are no $f_y(X, W, U_y)$ nor $f_y(X, Z, U_y)$ terms in $f_y$, we can write
\begin{align}
    &Y_{x', W_x}(u) = f_y^{(1)}(x', u_y) + f_y^{(2)}(W_{x}(u), Z(u), u_y) \\
    &\implies Y_{x', W_x}(u) - Y_{x, W_x}(u) = f_y^{(1)}(x', u_y) - f_y^{(1)}(x, u_y).
\end{align}
Since $U_y \ci X$, the above expression implies that $(Y_{x', W_x} - Y_{x, W_x}) \ci X$, which in turn implies that $x\text{-DE-SE}_{x_0, x_1}(y) = 0$. 
For the case in which there is a term $f_y(X, Z, U_y)$, we can again use the SCM $\mathcal{M}$ from Eqs.~\ref{eq:te-se-scm-1}-\ref{eq:te-se-scm-3}, in which the interaction exists by definition. 
For the case in which there is a term $f_y(X, W, U_y)$, consider the SCM $\mathcal{M}$ in Eqs.~\ref{eq:ex-de-se-1}-\ref{eq:ex-de-se-4} (from Ex.~\ref{ex:de-se-ia}) in which the interaction also exists by definitions. For both of these SCMs, we can also compute that $x\text{-DE-SE}_{x_0, x_1}(y) \neq 0$.

\paragraph{IE-SE Interaction.} We follow Fig.~\ref{fig:str-ie-se-flow} to provide the proof. Suppose that there is no $Z \to X$ edge. As before, due to the back-door criterion, we can show that $Y_{x', W_x} \ci X$, which implies that
\begin{align}
    &\ex[Y_{x_0, W_{x_1}} \mid X = x_1] = \ex[Y_{x_0, W_{x_1}} \mid X = x_0] \\
    &\ex[Y_{x_0, W_{x_0}} \mid X = x_1] = \ex[Y_{x_0, W_{x_0}} \mid X = x_0] \\
    &\qquad\qquad \implies x\text{-IE-SE}_{x_0, x_1}(y) = 0.
\end{align}
Further, consider the case in which there is no $X \to W$ edge. Therefore,
\begin{align}
    W_x (u) = W_{x'}(u) \implies Y_{x, W_{x'}} - Y_{x, W_{x}} = 0,
\end{align}
showing that $\ex[Y_{x, W_{x'}} - Y_{x, W_{x}} \mid x_1] = \ex[Y_{x, W_{x'}} - Y_{x, W_{x}} \mid x_0] = 0$, implying that $x\text{-IE-SE}_{x_0, x_1}(y) = 0$.

For the remainder, suppose that $Z \to X$ and $X \to W$ edges exist. We first consider the case in which there is no $f_w(X, Z, U_w)$ term. If there is term $f_y(Z, W, U_y)$, the interaction exists by definition. For an example, consider 
\begin{empheq}[left={\mathcal{M} = \empheqlbrace}]{align} 
  Z &\gets \text{Bernoulli}(0.5) \label{eq:ie-se-scm-1}\\
  X &\gets \text{Bernoulli}(0.5 + 0.2Z) \\
  W &\gets X + Z \\
  Y &\gets ZW, \label{eq:ie-se-scm-4}
\end{empheq}
in which we can also compute that $x\text{-IE-SE}_{x_0, x_1}(y) \neq 0$. If there is no term $f_y(Z, W, U_y)$ in $f_y$, then we further check if the $f_y(X, W, U_y)$ term in $f_y$ is linear in $W$, namely that
\begin{align}
    f_y(X, Z, W, U_y) = f_y^{(1)}(X, Z, U_y) + W f_y^{(2)}(X, U_y).
\end{align}
This allows us to write
\begin{align}
    Y_{x, W_{x'}}(u) = f_y^{(1)}(x, Z(u), u_y) + W_{x'}(u) f_y^{(2)}(x, u_y)
\end{align}
and, therefore, we have that
\begin{align}
    &Y_{x, W_{x'}}(u)  - Y_{x, W_{x}}(u) \\ &= [W_{x'}(u) - W_{x}(u)] f_y^{(2)}(x, u_y) \\
    &= [f_w^{(1)}(x', u_w) + f_w^{(2)}(Z(u), u_w) - (f_w^{(1)}(x, u_w) + f_w^{(2)}(Z(u), u_w))] f_y^{(2)}(x, u_y) \label{eq:no-fw-xz} \\
    &= [f_w^{(1)}(x', u_w) - f_w^{(1)}(x, u_w)] f_y^{(2)}(x, u_y), \label{eq:u-ie-uw-uy}
\end{align}
where Eq.~\ref{eq:no-fw-xz} follows from $W_{x}(u) = f_w^{(1)}(x, u_w) + f_w^{(2)}(Z(u), u_w)$ which holds true in this subcase (recall that there is no $f_w(X, Z, U_w)$ term in $f_w$, which allows for the expansion). Crucially, from Eq.~\ref{eq:u-ie-uw-uy} we can see that $Y_{x, W_{x'}}  - Y_{x, W_{x}} \ci X$, since $U_w, U_y \ci X$. From this, it follows that $x\text{-IE-SE}_{x_0, x_1}(y) = 0$. 
The requirement of linearity of $f_y$ in terms of $W$ for the non-existence of an interaction is motivated the following example:
\begin{empheq}[left={\mathcal{M} = \empheqlbrace}]{align} 
  Z &\gets \text{Bernoulli}(0.5) \label{eq:ie-se-scm-B-1}\\
  X &\gets \text{Bernoulli}(0.5 + 0.2Z) \\
  W &\gets X + Z \\
  Y &\gets W^2. \label{eq:ie-se-scm-B-4}
\end{empheq}
In $\mathcal{M}$, the paths $X \gets Z \to W$ and $X \to W$ do not interact at $W$ (since the function $f_w$ is additive in $X, Z$). However, spurious and indirect paths do interact at $Y$, since $Y$ is a non-linear function of $W$. This example demonstrates that interactions may exhibit non-Markovian behavior, in the sense that a non-interaction at $W$ induces an interaction at $Y$, even though $W$ is the only parent of $Y$.
In this SCM $\mathcal{M}$, we can also compute that $x\text{-IE-SE}_{x_0, x_1}(y) \neq 0$.

The final case we consider is when the $f_w(X, Z, U_w)$ term does exist in $f_w$. In this branch of the decision tree from Fig.~\ref{fig:str-ie-se-flow}, we check whether there is a $W \to Y$ edge. If not, we have that 
\begin{align}
    Y_{x, W_{x'}}(u) = Y_{x}(u)\; \forall u \implies Y_{x, W_{x'}}(u)  - Y_{x, W_{x}}(u) = 0 \;\forall u.
\end{align}
Therefore, any indirect effect equals $0$, and consequently also $x\text{-IE-SE}_{x_0, x_1}(y) = 0$. Finally, if a $W \to Y$ edge does exist, then the interaction exists by definition. One such example is the SCM
\begin{empheq}[left={\mathcal{M} = \empheqlbrace}]{align} 
  Z &\gets \text{Bernoulli}(0.5) \label{eq:ie-se-scm-C-1}\\
  X &\gets \text{Bernoulli}(0.5 + 0.2Z) \\
  W &\gets XZ \\
  Y &\gets W. \label{eq:ie-se-scm-C-4}
\end{empheq}
In the above SCM $\mathcal{M}$, we also have that $x\text{-IE-SE}_{x_0, x_1}(y) \neq 0$.

\paragraph{DE-IE-SE Interaction.} We consider Fig.~\ref{fig:str-de-ie-se-flow}. As before, if there is no $Z \to X$ edge, $Y_{x, W_{x'}} \ci X$, and hence also $x\text{-DE-IE-SE}_{x_0, x_1}(y) = 0$. If there is no $X \to W$ edge, then all indirect effects are 0, and hence $x\text{-DE-IE-SE}_{x_0, x_1}(y) = 0$. 

Consider now the case in which there is no $f_w(X, Z, U_w)$ term in $f_w$. We first check whether there exists a $f_y(X, Z, W, U_y)$ term in $f_y$. If there is no $f_y(X, Z, W, U_y)$ term, we further check if the $f_y(X, W, U_y)$ term in $f_y$ is linear in $W$. If yes, note that we can write
\begin{align}
    f_y(X, Z, W, U_y) = f_y^{(1)}(X, Z, U_y) + W f_y^{(2)}(X, U_y) + f_y^{(3)}(Z, W, U_y).
\end{align}
which allows us to express the potential outcome $Y_{x, W_{x'}}(u)$ as
\begin{align}
    Y_{x, W_{x'}}(u) = f_y^{(1)}(x, Z(u), u_y) + W_{x'}(u) f_y^{(2)}(x, u_y) + f_y^{(3)}(Z(u), W_{x'}(u), u_y).
\end{align}
Using this expression, we have that
\begin{align}
    &[Y_{x, W_{x'}}(u)  - Y_{x, W_{x}}(u)] - [Y_{x', W_{x'}}(u)  - Y_{x', W_{x}}(u)] \\
    &= [W_{x'}(u) - W_{x}(u)] [f_y^{(2)}(x, u_y) - f_y^{(2)}(x', u_y)] 
    \\&\quad + [f_y^{(4)}(Z(u), W_{x'}(u), u_y) - f_y^{(4)}(Z(u), W_{x}(u), u_y)] 
    \\&\quad - [f_y^{(4)}(Z(u), W_{x'}(u), u_y) - f_y^{(4)}(Z(u), W_{x}(u), u_y)] \\
    &= [f_w^{(1)}(x', u_w) + f_w^{(2)}(Z(u), u_w) - (f_w^{(1)}(x, u_w) + f_w^{(2)}(Z(u), u_w))] [f_y^{(2)}(x, u_y) - f_y^{(2)}(x', u_y)] \label{eq:fw-xz-expand}  \\
    &= [f_w^{(1)}(x', u_w) - f_w^{(1)}(x, u_w)] [f_y^{(2)}(x, u_y) - f_y^{(2)}(x', u_y)]  \label{eq:u-de-ie-uw-uy},
\end{align}
where Eq.~\ref{eq:fw-xz-expand} follows from $W_{x}(u) = f_w^{(1)}(x, u_w) + f_w^{(2)}(Z(u), u_w)$ which holds true in this subcase (recall that there is no $f_w(X, Z, U_w)$ term in $f_w$, which allows for this expansion). 
Based on Eq.~\ref{eq:u-de-ie-uw-uy}, we have that
\begin{align}
    \left( [Y_{x, W_{x'}}  - Y_{x, W_{x}}] - [Y_{x', W_{x'}}  - Y_{x', W_{x}}] \right) \ci X
\end{align}
since $U_w, U_y \ci X$. Therefore, we have that
\begin{align}
    &\ex \left( [Y_{x, W_{x'}}  - Y_{x, W_{x}}] - [Y_{x', W_{x'}}  - Y_{x', W_{x}}] \mid X = x_0 \right) \\ &= \ex \left( [Y_{x, W_{x'}}  - Y_{x, W_{x}}] - [Y_{x', W_{x'}}  - Y_{x', W_{x}}] \mid X = x_1 \right)
\end{align}
from which it follows that $x\text{-DE-IE-SE}_{x_0, x_1}(y) = 0$.
If the $f_y(X, W, U_y)$ term in $f_y$ is not linear in $W$, the interaction exists by definition. For an example, consider the following SCM:
\begin{empheq}[left={\mathcal{M} = \empheqlbrace}]{align} 
  Z &\gets \text{Bernoulli}(0.5) \label{eq:de-ie-se-scm-B-1}\\
  X &\gets \text{Bernoulli}(0.5 + 0.2Z) \\
  W &\gets X+Z \\
  Y &\gets XW^2. \label{eq:de-ie-se-scm-B-4}
\end{empheq}
In this SCM $\mathcal{M}$, we can also compute that $x\text{-DE-IE-SE}_{x_0, x_1}(y) \neq 0$.
Going back, we next consider the case in which $f_w$ does not have a term $f_w(X, Z, U_w)$ but $f_y$ does have a term $f_y(X, Z, W, U_y)$, in which case the interaction exists by definition. One such SCM can be given by: 
\begin{empheq}[left={\mathcal{M} = \empheqlbrace}]{align} 
  Z &\gets \text{Bernoulli}(0.5) \label{eq:de-ie-se-scm-A-1}\\
  X &\gets \text{Bernoulli}(0.5 + 0.2Z) \\
  W &\gets X+Z \\
  Y &\gets XZW. \label{eq:de-ie-se-scm-A-4}
\end{empheq}
Again, in the above SCM, we have that $x\text{-DE-IE-SE}_{x_0, x_1}(y) \neq 0$. 

The final case we consider is when there is a $f_w(X, Z, U_w)$ term in $f_w$. Then, if there is no edge $W \to Y$, there is no interaction, and we can easily show that $x\text{-DE-IE-SE}_{x_0, x_1}(y) = 0$ since all indirect effects are $0$, and $x\text{-DE-IE-SE}_{x_0, x_1}(y)$ is a linear combination of indirect effects. If there is an edge $W \to Y$, then an interaction does exist by definition, and for an example we can consider the SCM:
\begin{empheq}[left={\mathcal{M} = \empheqlbrace}]{align} 
  Z &\gets \text{Bernoulli}(0.5) \label{eq:de-ie-se-scm-C-1}\\
  X &\gets \text{Bernoulli}(0.5 + 0.2Z) \\
  W &\gets XZ \\
  Y &\gets XW. \label{eq:de-ie-se-scm-C-4}
\end{empheq}
In the above $\mathcal{M}$, we can again compute that $x\text{-DE-IE-SE}_{x_0, x_1}(y) \neq 0$.
\end{proof}
\section{{Experimental Details}} \label{appendix:exp-details}
\begin{figure}
    \centering
    \includegraphics[width=\linewidth]{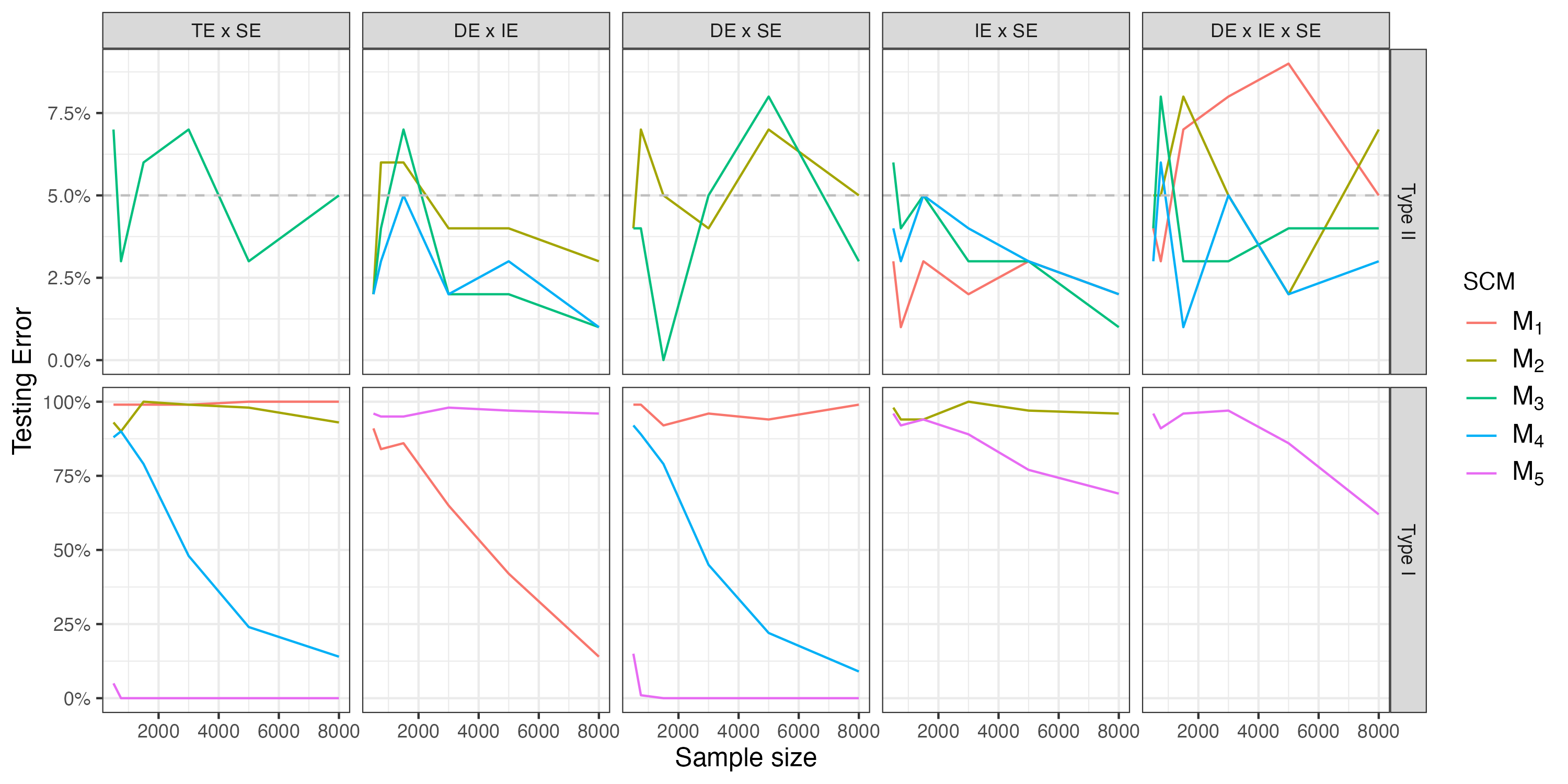}
    \caption{Type I and Type II errors for synthetic interaction tests.}
    \label{fig:alpha-testing}
\end{figure}

The SCMs $M_1$ to $M_5$ used in the synthetic experiments are given by the following sets of structural equations: 
{\allowdisplaybreaks
\begin{align}
%
%
\mathcal{M}_1 &= \left\{
\begin{aligned}
Z &\gets (Z_1, Z_2, Z_3) \sim N(0, I_3) \\
X &\gets \text{Bern}\left( \text{logit}(0.3 Z_1 - 0.2 Z_2 + 0.5 Z_3 + 0.2 Z_1^2) \right) \\
W &\gets \begin{pmatrix}
0.4 Z_1 + 0.1 Z_2 - 0.3 Z_3 + 0.5 X + \epsilon_1 \\
0.2 Z_1 - 0.1 Z_2 + 0.3 Z_3 + 0.4 X + \epsilon_2 \\
0.3 Z_1 - 0.2 Z_2 + 0.1 Z_3 + 0.3 X + \epsilon_3
\end{pmatrix}, \quad \epsilon_i \sim N(0, 1) \\
Y &\gets 0.5 W_1 + 0.4 W_2 + 0.3 W_3 + 0.2 Z_1 + 0.1 Z_2 + 0.4 Z_3 + 0.7 X \\&\quad + 0.2 X W_1 + \eta, \quad \eta \sim N(0, 1)
\end{aligned} \color{white} \right\} \\
%
%
\mathcal{M}_2 &= \left\{
\begin{aligned}
Z &\gets (Z_1, Z_2, Z_3) \sim (\text{Exp}(1), N(5, 1), \text{Uniform}(-2, 2)) \\
X &\gets \text{Bern}\left( \text{logit}(0.3 Z_1 - 0.2 Z_2 + 0.5 Z_3 + 0.2 Z_1^2) \right) \\
W &\gets \begin{pmatrix}
0.3 Z_1 - 0.5 Z_2 + 0.2 Z_3 + 0.2 X + \epsilon_1 \\
-0.1 Z_1 + 0.3 Z_2 + 0.1 Z_3 + 0.1 X + \epsilon_2 \\
0.2 Z_1 + 0.2 Z_2 - 0.3 Z_3 + 0.4 X + \epsilon_3
\end{pmatrix}, \quad \epsilon_i \sim N(0, 1) \\
Y &\gets 0.4 W_1 + 0.3 W_2 + 0.2 W_3 + 0.1 Z_1 + 0.3 Z_2 + 0.2 Z_3 + 0.4 X \\&\quad + (0.1W_1 - 0.3W_2 - 0.3W_3) (0.1Z_1 - 0.2Z_2 + 0.2 Z_3) + \eta, \quad \eta \sim N(0, 1)
\end{aligned} \color{white} \right\} \\
%
%
\mathcal{M}_3 &= \left\{
\begin{aligned}
Z &\gets (Z_1, Z_2, Z_3) \sim N(0, I_3) \\
X &\gets \text{Bern}\left( \text{logit}(0.3 Z_1 - 0.2 Z_2 + 0.5 Z_3 + 0.2 Z_1^2) \right) \\
W &\gets \begin{pmatrix}
0.3 Z_1^2 + \epsilon_1 \\
0.5 Z_2 + \epsilon_2 \\
0.4 X + \epsilon_3
\end{pmatrix}, \quad \epsilon_i \sim N(0, 1) \\
Y &\gets 0.3 W_1 + 0.2 W_2 + 0.1 W_3 + 0.2 Z_1 + 0.1 Z_2 + 0.3 Z_3 + \eta, \quad \eta \sim N(0, 1)
\end{aligned} \color{white} \right\} \\
%
%
\mathcal{M}_4 &= \left\{
\begin{aligned}
Z &\gets (Z_1, Z_2, Z_3) \sim N(0, I_3) \\
X &\gets \text{Bern}\left( \text{logit}(0.3 Z_1 - 0.2 Z_2 + 0.5 Z_3 + 0.2 Z_1^2) \right) \\
W &\gets \begin{pmatrix}
0.4 Z_1 + 0.1 Z_2 - 0.3 Z_3 + 0.5 X + \epsilon_1 \\
0.2 Z_1 - 0.1 Z_2 + 0.3 Z_3 + 0.4 X + \epsilon_2 \\
0.3 Z_1 - 0.2 Z_2 + 0.1 Z_3 + 0.3 X + \epsilon_3
\end{pmatrix}, \quad \epsilon_i \sim N(0, 1) \\
Y &\gets 0.5 W_1 + 0.4 W_2 + 0.3 W_3 + 0.2 Z_1 + 0.1 Z_2 + 0.4 Z_3 \\ &\quad + 0.3 X Z_1 + \eta, \quad  \eta \sim N(0, 1)
\end{aligned} \color{white} \right\} \\
%
%
\mathcal{M}_5 &= \left\{
\begin{aligned}
Z &\gets (Z_1, Z_2, Z_3) \sim N(0, I_3) \\
X &\gets \text{Bern}\left( \text{logit}(0.3 Z_1 - 0.2 Z_2 + 0.5 Z_3 + 0.2 Z_1^2) \right) \\
W &\gets \begin{pmatrix}
0.3 Z_1^2 + \epsilon_1 \\
0.5 Z_2 + \epsilon_2 \\
0.4 X + \epsilon_3
\end{pmatrix}, \quad \epsilon_i \sim N(0, 1) \\
Y &\gets 0.4 W_1 + 0.3 W_2 + 0.2 W_3 + 0.2 Z_1 + 0.1 Z_2 + 0.3 Z_3 \\
&\quad + 0.5 X Z_1 W_3 - 0.4 Z_2 W_3 + X Z_3 + \eta, \quad \eta \sim N(0, 1)
\end{aligned} \color{white} \right\}.
\end{align}}
\hspace*{-5pt}The summary of interactions is given in Tab.~\ref{tab:synthetic-ias}, and these results can be obtained by evaluating Defs.~\ref{def:str-ia} and \ref{def:str-ia-granular}. In the main text, we discussed the distribution of p-values for the hypothesis tests. Additionally, we know look at the Type I (false positive) and Type II (false negative) errors at the fixed significance level of $\alpha = 0.05$, shown in Fig.~\ref{fig:alpha-testing}. In the top row of the figure, corresponding to SCMs and interactions where the null-hypothesis of no interaction is true, we plot the Type II error rates. Note that these errors approximately fluctuate around 5\%, in line with the expected significance level. In the bottom row of the figure, corresponding to SCMs and interactions where the null-hypothesis of no interaction is false, we plot the Type I error rates. Visibly, for some interactions, with increased sample size we also have a decrease in the Type I error (or, in other words, an increase in power). For some interactions, however, the test statistic may be very close or equal to $0$ even if an interaction is present (see Sec.~\ref{sec:population-axis}), yielding a more difficult testing problem.

\section{Identification \& Estimation of Causal Quantities} \label{appendix:id-est}
In this appendix, we discuss the practical aspects of estimation of the causal quantities discussed in the main text.

\subsection{Identification of Causal Quantities}
Note that the quantities appearing in the TV decomposition in Eqs.~\ref{eq:tv-first-order}-\ref{eq:tv-third-order} are linear combinations (with coefficients in $\{-1, 1\}$) of potential outcomes of the form
\begin{align} \label{eq:po-xzw}
    \text{PO}(x_y, x_w, x_z) = \ex[ Y _{x_y, W_{x_w}} \mid X = x_z], 
\end{align}
where $x_y, x_w, x_z$ arbitrary combinations of values of $X$.
Based on the causal assumptions encoded in the causal diagram in Fig.~\ref{fig:sfm}, we first derive the identification expression for the potential outcome $\text{PO}(x_y, x_w, x_z)$.

\begin{figure}
    \centering
    \begin{tikzpicture}
	 [>=stealth, rv/.style={thick}, rvc/.style={triangle, draw, thick, minimum size=7mm}, node distance=18mm]
	 \pgfsetarrows{latex-latex};
	 \begin{scope}
		\node[rv] (Z) at (0,1) {$Z$};
	 	\node[rv] (X) at (-1.5,0.5) {$X = x_z$};
	 	\node[rv] (W) at (0,-0.75) {$W$};

            \node[rv] (Tx1w) at (1.75,-0.25) {$Y_{x_y, w}$};
            \node[rv] (Yxzw) at (1.75,0.75) {$Y_{x_y, z, w}$};
            \node[rv] (Wx0) at (0,-1.75) {$W_{x_w}$};
            \node[rv] (x1) at (-1.5,-0.25) {${x_y}$};
            \node[rv] (x0) at (-1.5,-1) {${x_w}$};
            \node[rv] (w) at (0.75,-1) {${w}$};
            \node[rv] (z) at (0.75,1.25) {${z}$};
	 	
            \draw[->] (Z) -- (W);
            \draw[->] (X) -- (W);
            \draw[->] (Z) -- (W);
            \draw[->] (x1) -- (Tx1w);
            \draw[->] (w) -- (Tx1w);
            \draw[->] (w) edge[bend left = 20] (Yxzw);
            \draw[->] (z) edge[bend left = 20] (Yxzw);
            \draw[->] (x1) edge[bend left = 0] (Yxzw);
            \draw[->] (x0) -- (Wx0);
            \draw[->] (Z) -- (Tx1w);
            \draw[->] (Z) edge[bend left = -30] (Wx0);
		\path[<->] (Z) edge[bend left = -30, dashed] (X);
            \path[<->] (W) edge[bend left = 30, dashed] (Wx0);
	 \end{scope}
	 \end{tikzpicture}
    \caption{Counterfactual graph of the causal graph in Fig.~\ref{fig:ctf-sfm}.}
    \label{fig:ctf-sfm}
\end{figure}
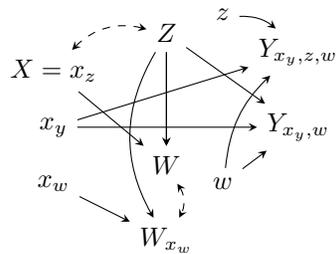

{\allowdisplaybreaks
\begin{align}
        \ex[Y_{x_y, W_{x_w}} \mid x_z] &= \sum_z \ex[Y_{x_y, W_{x_w}} \mid x_z, z] \pr(z \mid x_z) \\
        &= \sum_{z, w} \ex[Y_{x_y, w, z} \mathbb{1}(W_{x_w} = w) \mid x_z, z] \pr(z \mid x_z) \\
        &= \sum_{z, w} \ex[Y_{x_y, w, z} \mid x_z, z] \ex[\mathbb{1}(W_{x_w} = w) \mid x_z, z] \pr(z \mid x_z) \label{eq:s-w-indep}\\
        &= \sum_{z, w} \ex[Y_{x_y, w, z}] \pr(W_{x_w} = w \mid x_z, z) \pr(z \mid x_z) \\
        &= \sum_{z, w} \ex[Y \mid {x_y, w, z}] \pr(W_{x_w} = w \mid x_z, z) \pr(z \mid x_z) \\
        &= \sum_{z, w} \ex[Y | {x_y, w, z}] \pr(W = w \mid x_w, z) \pr(z \mid x_z). \label{eq:po-xzw-id}
\end{align}}
The first line follows by the law of total probability, the second by counterfactual unnesting of $W_{x_w}$, the third by the independence $Y_{x_y, z,w} \ci W_{x_w} \mid X, Z$ implied by the counterfactual graph in Fig.~\ref{fig:ctf-sfm}, the fourth by the independence $Y_{x_y, z,w} \mid X, Z$, the fifth by the same independence and the consistency axiom, and the sixth by the independence $W_{x_w} \ci X \mid Z$ and the consistency axiom. The expression in Eq.~\ref{eq:po-xzw-id} is the identification expression for the potential outcome $\text{PO}(x_y, x_w, x_z)$.

\subsection{Influence Functions}
We next discuss how to estimate the $\text{PO}(x_y, x_w, x_z)$ using a finite dataset. To perform efficient estimation, we derived the influence function for the estimating expression in Eq.~\ref{eq:po-xzw-id}. For this, we follow the heuristics from \citep{kennedy2022semiparametric} to derive the expression for the influence function, but omit formal proofs that this influence function satisfies Neyman orthogonality. We denote the identification expression in Eq.~\ref{eq:po-xzw-id} with $\psi$, and derive the influence function as follows:
\begin{align}
    \ifc(\psi) &= \left\{ \sum_{z, w} \ex[Y | {x_y, w, z}] \pr(w \mid x_w, z) \pr(z \mid x_z)  \right\} \\
    &= \sum_{z, w} \underbrace{\ifc (\ex[Y | {x_y, w, z}]) \pr(w \mid x_w, z) \pr(z \mid x_z)}_{T_1} \\
    &\qquad + \underbrace{\ex[Y | {x_y, w, z}] \ifc ( \pr(w \mid x_w, z) ) \pr(z \mid x_z)}_{T_2} \\
    &\qquad + \underbrace{\ex[Y | {x_y, w, z}] \pr(w \mid x_w, z) \ifc ( \pr(z \mid x_z) )}_{T_3}.
\end{align}
We have that
\begin{align}
    \ifc (\ex[Y | {x_y, w, z}]) &= \frac{\mathbbm{1}(X = x_y, Z = z, W = w)}{\pr(x_y, z, w)} [Y - \ex[Y \mid x_y, z, w]] \label{eq:if-t1} \\
    \ifc ( \pr(w \mid x_w, z) ) &= \frac{\mathbbm{1}(X = x_w, Z = z)}{\pr(x_w, z)} [\mathbbm{1}(W = w) - \pr(w \mid x_w, z)] \\
    \ifc ( \pr(z \mid x_z) ) &= \frac{\mathbbm{1}(X = x_z)}{\pr(x_z)} [\mathbbm{1}(Z = z) - \pr(z \mid x_z)]. \label{eq:if-t3}
\end{align}
Therefore, we have that
\begin{align}
    T_1 &= \frac{\mathbbm{1}(X = x_y, Z = z, W = w)}{\pr(x_y, z, w)} [Y - \ex[Y \mid x_y, z, w]] \pr(w \mid x_w, z) \pr(z \mid x_z) \\
    &= \frac{\pr(w \mid x_w, z) \pr(z \mid x_z)}{\pr(x_y, z, w)}  \mathbbm{1}(X = x_y, Z = z, W = w) [Y - \ex[Y \mid x_y, z, w]] \\
    &= \frac{\pr(w, x_w, z) \pr(z, x_z)}{\pr(x_w, z) \pr(x_z) \pr(x_y, z, w)} \mathbbm{1}(X = x_y, Z = z, W = w) [Y - \ex[Y \mid x_y, z, w]] \\
    &= \frac{1}{\pr(x_z)} \frac{\pr(x_z \mid z)}{\pr(x_w \mid z)} \frac{\pr(x_w \mid z, w)}{\pr(x_y \mid z, w)} \mathbbm{1}(X = x_y, Z = z, W = w) [Y - \ex[Y \mid x_y, z, w]].
\end{align}
We further have that
\begin{align}
    T_2 &= \ex[Y | {x_y, w, z}] \frac{\mathbbm{1}(X = x_w, Z = z)}{\pr(x_w, z)} [\mathbbm{1}(W = w) - \pr(w \mid x_w, z)] \pr(z \mid x_z) \\
    &= \frac{\pr(x_z \mid z)}{\pr(x_w \mid z)} \frac{\mathbbm{1}(X = x_w, Z = z)}{\pr(x_z)}  [ \ex[Y | {x_y, w, z}] - \ex[Y | {x_y, w, z}] \pr(w \mid x_w, z)  ], \\
    T_3 &= \ex[Y | {x_y, w, z}] \pr(w \mid x_w, z) \frac{\mathbbm{1}(X = x_z)}{\pr(x_z)} [\mathbbm{1}(Z = z) - \pr(z \mid x_z)] \\
    &= \frac{\mathbbm{1}(X = x_z, Z = z)}{\pr(x_z)} \ex[Y | {x_y, w, z}] \pr(w \mid x_w, z) \\&\quad -   \frac{\mathbbm{1}(X = x_z)}{\pr(x_z)} \ex[Y | {x_y, w, z}] \pr(w \mid x_w, z) \pr(z \mid x_z).
\end{align}
Plugging in the expressions for $T_1, T_2, T_3$ into the expression for $\ifc (\psi)$, we obtain:
\begin{align}
    \ifc(\psi) &= \frac{\mathbbm{1}(X = x_y)}{\pr(x_z)} \frac{\pr(x_z \mid Z)}{\pr(x_w \mid Z)} \frac{\pr(x_w \mid Z, W)}{\pr(x_y \mid Z, W)}  [Y - \mu(x_y, Z, W)] \\
    &\quad + \frac{\mathbbm{1}(X = x_w)}{\pr(x_z)} \frac{\pr(x_z \mid Z)}{\pr(x_w \mid Z)} [\mu(x_y, Z, W) - \ex [\mu(x_y, Z, W) \mid x_w, Z] ]  \\
    &\quad + \frac{\mathbbm{1}(X = x_z)}{\pr(x_z)} \ex [\mu(x_y, Z, W) \mid x_w, Z] - \psi.
\end{align}
Here, $\mu(x, z, w) = \ex[Y \mid X = x, Z = z, W = w]$, and $\ex [\mu(x_y, Z, W) \mid x_w, Z]$ is the nested mean
\begin{align}
    \sum_w \ex[Y \mid X = x_y, Z, W=w] \pr(W = w \mid X = x_w, Z).
\end{align}
Analogously to the above derivation, we have that the identification expression for the log-risk scale potential outcome is given by
\begin{align}
    \ex[\log P_{x_y, W_{x_w}} \mid x_z]= \sum_{z, w} \log \pr(Y = 1 \mid {x_y, w, z}) \pr(w \mid x_w, z) \pr(z \mid x_z). \label{eq:po-lr-xzw-id}
\end{align}
Therefore, deriving its influence function follows very similar steps, with the difference that
\begin{align}
    \ifc ( \log \pr(Y = 1 \mid {x_y, w, z}) ) = \frac{\mathbbm{1}(X = x_y, Z = z, W = w)}{\pr(Y = 1 \mid {x_y, w, z}) \pr(x_y, z, w)}[Y - \pr(Y = 1 \mid {x_y, w, z})],
\end{align}
obtained through the chain rule $\ifc (f(\psi)) = f'(\psi) \ifc (\psi)$, and using the known influence function of $\pr(y \mid {x_y, w, z})$. Of course, for a binary outcome $Y$ considered, $\pr(Y = 1 \mid {x_y, w, z})$ is also equal to $\mu(x_y, z, w)$, and we can therefore write the influence function for the identification expression $\psi_{lr}$ on the RHS of Eq.~\ref{eq:po-lr-xzw-id} as:
\begin{align}
    \ifc(\psi_{lr}) &= 
    \frac{\mathbbm{1}(X = x_y)}{\pr(x_z)} \frac{\pr(x_z \mid Z)}{\pr(x_w \mid Z)} \frac{\pr(x_w \mid Z, W)}{\pr(x_y \mid Z, W)}  \frac{[Y - \mu(x_y, Z, W)]}{\mu(x_y, Z, W)} \\
    &\quad + \frac{\mathbbm{1}(X = x_w)}{\pr(x_z)} \frac{\pr(x_z \mid Z)}{\pr(x_w \mid Z)} [\log \mu(x_y, Z, W) - \ex [\log \mu(x_y, Z, W) \mid x_w, Z] ]  \\
    &\quad + \frac{\mathbbm{1}(X = x_z)}{\pr(x_z)} \ex [\log \mu(x_y, Z, W) \mid x_w, Z] - \psi_{lr}.
\end{align}
\section{Log-Risk and Log-Odds Scales} \label{appendix:lr-lo-scales}
\newcommand{\rrnm}{LR}
\newcommand{\ornm}{LO}
The main text of the paper is focused on interaction testing on the difference scale, where two potential outcomes $Y_{C_1}, Y_{C_0}$ were compared by taking the expected difference
\begin{align}
    \ex[Y_{C_1} - Y_{C_0}].
\end{align}
Here, the outcome $Y$ is considered to be continuous. Further, we also discussed how to perform interaction testing for a binary outcome $Y$, in which case the potential outcome $Y_C$ is replaced by $P_C$, where the random variable $P$ is the risk, defined through $P_C = \ex_{U_y} Y_C$. In such a setting, we would talk about the \textit{risk difference} scale, and would use the measure $\ex[P_{C_1} - P_{C_0}]$, which in fact equals $\ex[Y_{C_1} - Y_{C_0}]$, since $\ex \; \ex_{U_y} Y_C = \ex\, Y_C$. Therefore, on the difference and risk difference scales, essentially the same measures for quantifying interactions are used, although the structural notions of interaction are slightly adapted (recall Def.~\ref{def:str-ia} and Rem.~\ref{rem:binary-y}). 

Oftentimes, especially when dealing with binary outcomes, different scales may be interesting, such as the log-risk difference scale or the log-odds difference scale (to abbreviate, we talk about the log-risk scale or LR scale, and similarly for the log-odds or LO scale). 
In this appendix, we describe how the results of the manuscript can be extended to such alternative scales, with a particular focus on binary outcomes.

For the case of binary outcomes, as discussed in Rem.~\ref{rem:binary-y}, the key was to replace the structural mechanism $f_y$ with its probability counterpart, defined by:
\begin{align}
    p_y(x, z, w) := \ex_{U_y}[Y_{x, z, w} ] = \pr_{U_y}(Y_{x, z, w} = 1),
\end{align}
The structural mechanism $f_y$ returns a binary value of $Y$, while $p_y$ returns a probability of $Y$ belonging to class $1$ given covariates $X=x, Z=z, W=w$.
When considering other scales apart from the difference scale, we may consider suitable transformations of the constructed mechanism $p_y(x, z, w)$. In fact, when the data corresponds to the causal diagram in Fig.~\ref{fig:sfm}, the outcome $Y$ can be thought of as being simply replaced by the random variable $P = \ex_{U_y} \, Y$, and the same causal diagram is still valid with $P$ replacing the $Y$ outcome. For the log-risk scale, we consider the transformation 
\begin{align}
    \log{p_y(x, z, w)},
\end{align}
whereas for the log-odds scale we consider the transformation
\begin{align}
    \text{logit}\,{p_y(x, z, w)},
\end{align}
where $\text{logit}(a) = \log{\frac{a}{1-a}}$. 
Analogously to the notion of no interaction on a difference scale in Def.~\ref{def:str-ia}, we can consider no interaction criteria on \rrnm{} and \ornm{} scales, which are closely related:
\begin{definition}[Structural Interaction Criteria for \rrnm{} and \ornm{} Scales] \label{def:str-ia-rr-or}
Consider the causal diagram in Fig.~\ref{fig:conf-graph}, and let $p_y(x, z)$ be the probability $P(Y = 1 \mid x, z)$. We say that
\begin{enumerate}[label=(\alph*)]
    \item There is no structural interaction of total and spurious effects on the log-risk scale, written Str-T\rrnm{}-S\rrnm{} = 0, if the mechanism $\log p_y(x,z)$ satisfies Str-TE-SE = 0. Explicitly, Str-T\rrnm{}-S\rrnm{} = 0 if either we can write:
    \begin{align}
        p_y(x, z) = \exp{ \{ f^{(1)}_y(x) + f^{(2)}_y(z) \} },
    \end{align}
    or if there is no back-door path between $X, P$.
    \item There is no total-spurious interaction on the odds ratio scale, written Str-T\ornm{}-S\ornm{} = 0, if the mechanism $\mathrm{logit}\; p_y(x,z)$ satisfies Str-TE-SE = 0. Explicitly, Str-T\ornm{}-S\ornm{} = 0 if either we can write
    \begin{align}
        p_y(x, z) = \mathrm{expit}\,\{ f^{(1)}_y(x) + f^{(2)}_y(z) \},
    \end{align}
    or if there is no back-door path between $X, P$. Here, $\mathrm{expit}(a) := \exp{(a)}/(1+\exp{(a)})$.
\end{enumerate}
Next, consider the causal diagram in Fig.~\ref{fig:sfm}, and let $p_y(x, z, w)$ be the probability $P(Y = 1 \mid x, z, w)$. We say that
\begin{enumerate}[label=(\alph*)]
    \addtocounter{enumi}{2}
    \item There is no direct-indirect structural interaction for the log-risk scale, written Str-D\rrnm{}-I\rrnm{} = 0, if $\log p_y(x, z, w)$ satisfies Str-DE-IE. Explicitly, Str-D\rrnm{}-I\rrnm{} = 0 if either we can write
    \begin{align}
        p_y(x, z, w) = \exp{ \{ f^{(1)}_y(x, z) + f^{(2)}_y(z, w) \} },
    \end{align}
    or if there is no indirect path $X \to W \to P$.

    \item There is no direct-indirect structural interaction for the odds ratio scale, written Str-D\ornm{}-I\ornm{} = 0, if $\mathrm{logit}\; p_y(x, z, w)$ satisfies Str-DE-IE. Explicitly, Str-D\ornm{}-I\ornm{} = 0 if either we can write
    \begin{align}
        p_y(x, z, w) = \mathrm{expit}\,{ \{ f^{(1)}_y(x, z) + f^{(2)}_y(z, w) \} },
    \end{align}
    or if there is no indirect path $X \to W \to P$.
\end{enumerate}
If interactions exist, we say that the corresponding structural criterion is equal to $1$.  
\end{definition}
The no interaction notions, on a structural level, are very similar (analogous) for the \rrnm{} and \ornm{} scales, and can be written using the structural interaction notions on the difference scale. The key subtlety is that we need to consider appropriate transformations of the probability of a positive outcome, $\log P(y \mid x, z, w)$ for the \rrnm{} scale, and $\mathrm{logit}\; P(y \mid x, z, w)$ for the \ornm{} scale. We remark that the more granular interactions (DE-SE, SE-IE, and DE-IE-SE) can also be written for \ornm{} and \rrnm{} scales, but we omit these definitions in the interest of brevity.

Naturally, the transformations that are used also have implications for how we can test for interactions, and how the TV measure is decomposed. We now discuss the log-risk scale.

\subsection{Log-Risk Scale} \label{appendix:lr-scale}
The next task at hand is to derive appropriate measures for testing structural interactions on the LR and LO scales. 
To derive such measures, we may be tempted to use known approaches in the literature which have considered the \textit{risk ratio} scale \citep{rothman2008modern, vanderweele2015explanation}. 
A common way of obtaining measures for the risk ratio scale is to replace the subtraction operator with the division operator when comparing contrasts. For instance, on the difference scale, we considered the measure of direct effect called $x$-DE$_{x_0, x_1}(y \mid x)$, defined as
\begin{align}
    \ex[Y_{x_1, W_{x_0}} - Y_{x_0, W_{x_0}} \mid X = x].
\end{align}
When replacing subtraction with division, we derive the measure which we call $x$-DRR$_{x_0, x_1}(y \mid x)$
\begin{align}
    \frac{\ex[Y_{x_1, W_{x_0}} \mid X = x]}{\ex[Y_{x_0, W_{x_0}} \mid X = x]}.
\end{align}
Notions of indirect or spurious effects, and also interaction effects, can be defined similarly. However, as the following example illustrates, more care needs to be taken, since the measures on the risk ratio scale are in fact not admissible for detecting interactions according to Def.~\ref{def:str-ia-rr-or}:
\begin{example}[Risk Ratio Measures' Non-Admissibility] \label{ex:rr-measure-failure}
    Consider the SCM $\mathcal{M}$ given by:
    \begin{empheq}[left={\mathcal{M} = \empheqlbrace}]{align} 
    Z &\gets \text{Bernoulli}(0.5) \label{eq:rr-fail-scm-1}\\
    X &\gets \text{Bernoulli}(0.5) \\
    W &\gets \text{Bernoulli}(0.5 + 0.1X) \\
    Y &\gets \text{Bernoulli}(\exp{(X + XZ + W - 3)}). \label{eq:rr-fail-scm-4}
\end{empheq}
To test for an interaction of direct and indirect pathways at the log-risk scale, we may consider the DRR-IRR quantity (analogous to the $x$-DE-IE quantity from Def.~\ref{def:xspecific}), and to test whether
\begin{align} \label{eq:rr-test-attempt}
    \text{DRR-IRR}_{x_0,x_1}(y) = \frac{\ex[Y_{x_1, W_{x_0}}]}{\ex[Y_{x_0, W_{x_0}}]} \times \frac{\ex[Y_{x_0, W_{x_1}}]}{\ex[Y_{x_1, W_{x_1}} ]} = 1.
\end{align}
The quantity tests whether the risk ratio of a $x_0 \to x_1$ transition is the same for the two settings in which $W$ varies naturally as $W_{x_0}$ vs. $W_{x_1}$.
Based on the SCM $\mathcal{M}$, we can compute that
\begin{align}
    \ex[Y_{x_1, W_{x_0}}] &= \frac{1}{4} ( 1 + 2e^{-1} + e^{-2}) \\
    \ex[Y_{x_0, W_{x_0}}] &= \frac{1}{4} ( 2e^{-3} + 2e^{-2}) \\
    \ex[Y_{x_0, W_{x_1}}] &= \frac{1}{5} ( 2e^{-3} + 3e^{-2}) \\
    \ex[Y_{x_1, W_{x_1}}] &= \frac{1}{10} (3 +  5e^{-1} + 2e^{-2}).
\end{align}
Therefore, we have that
\begin{align}
    \frac{\ex[Y_{x_1, W_{x_0}}]}{\ex[Y_{x_0, W_{x_0}}]} \times \frac{\ex[Y_{x_0, W_{x_1}}]}{\ex[Y_{x_1, W_{x_1}} ]} =
    \frac{1 + 2e^{-1} + e^{-2}}{2e^{-3} + 3e^{-2}} \times \frac{4e^{-3} + 6e^{-2}}{3 +  5e^{-1} + 2e^{-2}} \approx 0.73 \neq 1.
\end{align}
However, after a log-transformation, the mechanism of the probability of $Y$ is given by
\begin{align}
    \log p_y(X, Z, W) = X + XZ + W - 3,
\end{align}
and has no interaction term between $X, W$. Therefore, the measure on the LHS of Eq.~\ref{eq:rr-test-attempt} is not admissible with respect to the structural notion of interaction of pathways on the log-risk scale, labeled Str-D\rrnm{}-I\rrnm{}.
\end{example}
Based on the above example, we can see that the common approach for translating measures on a difference scale to measures on a risk ratio scale, by replacing subtraction with division \citep{rothman2008modern, vanderweele2015explanation}, is not sufficient for our purposes, and a different approach is needed. Motivated by this observation, we first introduce the log-transformed measures of direct, indirect, and spurious effects, and their relevant interactions:
\begin{definition}[$x$-specific T\rrnm{}, D\rrnm{}, I\rrnm{}, and S\rrnm{}] \label{def:x-specific-LRs}
The $x$-\{total, direct, indirect, spurious\} log-risk measures are defined as follows:
    \begin{align}
    x\text{-T\rrnm{}}_{x_0, x_1}(\log p \mid x) &= \ex\left[\log P_{x_1} \mid x\right] - \ex\left[\log P_{x_0} \mid x\right] \\
    x\text{-D\rrnm{}}_{x_0, x_1}(\log p \mid x) &= \ex\left[\log P_{x_1, W_{x_0}} \mid x\right] - \ex\left[\log P_{x_0} \mid x\right] \\
    x\text{-I\rrnm{}}_{x_0, x_1}(\log p \mid x) &= \ex\left[\log P_{x_0, W_{x_1}} \mid x\right] - \ex\left[\log P_{x_0} \mid x\right] \\
    x\text{-S\rrnm{}}_{x_0, x_1}(\log p) &= \ex\left[\log P_{x_0} \mid x_1\right] - \ex\left[\log P_{x_0} \mid x_0\right].
\end{align}
    The $x$-specific T\rrnm{}-S\rrnm{} and D\rrnm{}-I\rrnm{} interaction measures are defined as
    \begin{align}
        x\text{-D\rrnm{}-I\rrnm{}}_{x_0, x_1}(\log p \mid x) &= \ex\left[\log P_{x_1, W_{x_1}} - \log P_{x_0, W_{x_1}} \mid x\right] \\ &\quad - \ex\left[\log P_{x_1, W_{x_0}} - \log P_{x_0}\mid x\right] \\
        x\text{-T\rrnm{}-S\rrnm{}}_{x_0, x_1}(\log p) &= \ex\left[\log P_{x_1} - \log P_{x_0} \mid x_1\right] -
        \ex\left[\log P_{x_1} - \log P_{x_0} \mid x_0\right].
    \end{align}
\end{definition}
The interpretation of interaction measures at the \rrnm{} scale is related to the difference scale. The $x\text{-D\rrnm{}-I\rrnm{}}_{x_0, x_1}(\log p \mid x)$ looks at the increase in expected log-risk from a $x_0 \to x_1$ transition along the direct path while having $W_{x_1}$ for the group of units $X(u) = x$, compared to the increase in expected log-risk of the same transition while having $W_{x_0}$. In other words, the quantity is trying to compute how much changing $W_{x_0} \to W_{x_1}$ modifies the increase in the expected log-risk associated with a direct $x_0 \to x_1$ transition. 
Similarly, the $x\text{-T\rrnm{}-S\rrnm{}}_{x_0, x_1}(y)$ quantifies how much changing $X = x_0$ to $X = x_1$ along the spurious path modifies the change in expected log-risk of a $x_0 \to x_1$ transition along the causal paths. 

Armed with the $x$-specific measures for quantifying interaction of pathways on the log-risk scale (Def.~\ref{def:x-specific-LRs}), we can investigate the analogue of the TV decomposition, offering the formal result that mirrors Thm.~\ref{thm:1st-tv-decomp}:
\begin{theorem}[\rrnm{} Total Variation Decomposition] \label{thm:1st-tvr-decomp}
    Let the total variation log-risk measure be denoted by TV\rrnm{}$_{x_0, x_1}(\log p)$ and defined as $\ex[\log P \mid x_1] - \ex[\log P \mid x_1]$. The TV\rrnm{}$_{x_0, x_1}(\log p)$ can be decomposed as follows:
    \begin{align}
        \text{TV\rrnm{}}_{x_0, x_1}(\log p) = {x\text{-T\rrnm{}}_{x_0, x_1}(\log p\mid x_0)} + {x\text{-S\rrnm{}}_{x_0, x_1}(\log p)} + x\text{-T\rrnm{}-S\rrnm{}}_{x_0, x_1}(\log p).
    \end{align}
    The TV\rrnm{} measure can also be decomposed as:
    \begin{align} \label{eq:tvr-decomp-1st}
        \text{TV\rrnm{}}_{x_0, x_1}(\log p) &= {x\text{-D\rrnm{}}_{x_0, x_1}(\log p\mid x_0)} + {x\text{-I\rrnm{}}_{x_0, x_1}(\log p\mid x_0)} \\&\quad + {x\text{-D\rrnm{}-I\rrnm{}}_{x_0, x_1}(\log p\mid x_0)} \\
        &\quad + {x\text{-S\rrnm{}}_{x_0, x_1}(\log p)} + x\text{-T\rrnm{}-S\rrnm{}}_{x_0, x_1}(\log p). \nonumber
    \end{align}
\end{theorem}
Once again, the T\rrnm{}-S\rrnm{} and D\rrnm{}-I\rrnm{} interaction measures can be related to the structural level, through the following admissibility result:
\begin{proposition}[Structural Admissibility of \rrnm{} Interaction Tests] \label{prop:str-admissibility-rr}
    \begin{align}
        \text{Str-T\rrnm{}-S\rrnm{}} = 0 &\implies x\text{-T\rrnm{}-S\rrnm{}}_{x_0, x_1}(\log p) = 0, \\
        \text{Str-D\rrnm{}-I\rrnm{}} = 0 &\implies x\text{-D\rrnm{}-I\rrnm{}}_{x_0, x_1}(\log p \mid x) = 0.
    \end{align}
\end{proposition}
In words, whenever we can either (i) write the quantity $\log p_y(x, z)$ without a functional term $f(x, z)$ taking both $x, z$ as inputs or (ii) there is no back-door path between $X, P$, then the interaction $x\text{-T\rrnm{}-S\rrnm{}}_{x_0, x_1}(\log p)$ must equal $0$. Similarly, whenever we can either (iii) write $\log p_y(x, z, w)$ without a functional term $f(x, w)$ taking both $x, w$ as inputs or (iv) there is no $X \to W \to P$ indirect path, then the interaction measure $x\text{-D\rrnm{}-I\rrnm{}}_{x_0, x_1}(\log p \mid x)$ must equal $0$. The admissibility result from Prop.~\ref{prop:str-admissibility-rr} again leads to an interaction testing procedure, shown in Alg.~\ref{algo:ia-testing-rr}.
\begin{algorithm}[t]
    \caption{Interaction Testing for TV\rrnm{} Decomposition}
     \begin{algorithmic}[1]
        \Statex \textbullet~\textbf{Inputs:} Causal Diagram $\mathcal{G}$, Observational Data $\mathcal{D}$
        \Statex Compute the estimate of the total-spurious interaction log-risk $x\text{-T\rrnm{}-S\rrnm{}}_{x_0, x_1}(y)$, and its 95\% confidence interval. Test the hypothesis
        \begin{align}
            H^{\text{T\rrnm{}-S\rrnm{}}}_0: x\text{-T\rrnm{}-S\rrnm{}}_{x_0, x_1}(\log p) = 0.
        \end{align}
        \Statex Compute the estimate of the direct-indirect interaction log-risk $x\text{-D\rrnm{}-I\rrnm{}}_{x_0, x_1}(\log p \mid x_0)$, and its 95\% confidence interval. Test the hypothesis
        \begin{align}
            H^{\text{D\rrnm{}-I\rrnm{}}}_0: x\text{-D\rrnm{}-I\rrnm{}}_{x_0, x_1}(\log p \mid x_0) = 0.
        \end{align}
        \Statex \begin{itemize}
            \item if neither $H^{\text{T\rrnm{}-S\rrnm{}}}_0, H^{\text{D\rrnm{}-I\rrnm{}}}_0$ are rejected, return the decomposition
            \begin{align}
                \text{TV\rrnm{}}_{x_0, x_1}(\log p) &= {x\text{-D\rrnm{}}_{x_0, x_1}(\log p\mid x_0)} + {x\text{-I\rrnm{}}_{x_0, x_1}(\log p\mid x_0)} \\&\quad + {x\text{-S\rrnm{}}_{x_0, x_1}(\log p)}. \nonumber 
            \end{align}
            \item if only $H^{\text{D\rrnm{}-I\rrnm{}}}_0$ is rejected, return the decomposition
            \begin{align}
            \text{TV\rrnm{}}_{x_0, x_1}(\log p) &= {x\text{-D\rrnm{}}_{x_0, x_1}(\log p\mid x_0)} + {x\text{-I\rrnm{}}_{x_0, x_1}(\log p\mid x_0)}\\ &\quad +  x\text{-D\rrnm{}-I\rrnm{}}_{x_0, x_1}(\log p \mid x_0)
            + {x\text{-S\rrnm{}}_{x_0, x_1}(\log p)}. \nonumber
         \end{align}
         \item if only $H^{\text{T\rrnm{}-S\rrnm{}}}_0$ is rejected, return the decomposition
            \begin{align}
            \text{TV\rrnm{}}_{x_0, x_1}(\log p) &= {x\text{-D\rrnm{}}_{x_0, x_1}(\log p\mid x_0)} + {x\text{-I\rrnm{}}_{x_0, x_1}(\log p\mid x_0)} \\ &\quad
            + {x\text{-S\rrnm{}}_{x_0, x_1}(\log p)} + {x\text{-T\rrnm{}-S\rrnm{}}_{x_0, x_1}(\log p)}. \nonumber
         \end{align}
        \item if both $H^{\text{T\rrnm{}-S\rrnm{}}}_0, H^{\text{D\rrnm{}-I\rrnm{}}}_0$ are rejected, return the TV\rrnm{} decomposition in Eq.~\ref{eq:tvr-decomp-1st}.
        \end{itemize}
        \Statex \textbullet~\textbf{Output:} TV\rrnm{} decomposition with parsimony.
     \end{algorithmic}
     \label{algo:ia-testing-rr}
\end{algorithm}
For instance, the hypothesis $H^{\text{T\rrnm{}-S\rrnm{}}}_0$ tests whether
\begin{align}
    \ex\left[\log \frac{P_{x_1}}{P_{x_0}} \mid x_1\right] =
    \ex\left[\log \frac{P_{x_1}}{P_{x_0}} \mid x_0\right],
\end{align}
or in words, if the change in expected log-risk of a causal $x_0 \to x_1$ transition is equal for the $X(u) = x_1$ and $X(u) = x_0$ groups. The test can, by symmetry, also be understood as testing if
\begin{align}
    \ex\left[\log P_{x_1} \mid x_1\right] - \ex\left[\log P_{x_1} \mid x_0\right] = \ex\left[\log P_{x_0} \mid x_1\right] - \ex\left[\log P_{x_0} \mid x_0\right].
\end{align}
In words, the test is comparing the change in expected log-risk of a spurious $x_0 \to x_1$ transition while having $X = x_1$ along causal paths vs. the change in expected log-risk of the same transition while having $X = x_0$ along causal paths. Naturally, the $H^{\text{D\rrnm{}-I\rrnm{}}}_0$ hypothesis is amenable to similar interpretations.

\paragraph{Odds Ratio Scale.} When considering the log-risk scale measures from Def.~\ref{def:x-specific-LRs}, we implicitly replaced the potential outcome $Y_C$ (used on the difference scale) with the logarithm of its probability of being positive $\log P_C = \log \ex_{U_y} Y_C$, where $\ex_{U_y}$ denotes that the randomness in the $U_y$ noise variable is integrated out. When considering the log-odds scale, we deploy a very similar approach, and instead use the log-odds transformation, and the potential outcome $\log O_C$, given by
\begin{align}
    \log O_C = \mathrm{logit}\, P_C = \log \frac{\ex_{U_y} Y_C}{1 - \ex_{U_y} Y_C}.
\end{align}
Using this notion, we can again write the definitions of $x$-specific log-odds scale effects, analogously to Def.~\ref{def:x-specific-LRs}. For instance, the $x$-specific direct log-odds effect would be defined as 
\begin{align}
    x\text{-D\ornm{}}_{x_0, x_1}(\log o \mid x) &= \ex\left[\log O_{x_1, W_{x_0}} \mid x\right] - \ex\left[\log O_{x_0} \mid x\right],
\end{align}
and all other effects would be defined similarly, with $O_C$ replacing $P_C$ in the expressions in Def.~\ref{def:x-specific-LRs}. Importantly, the log-odds scale effects would (i) satisfy admissibility with respect to the structural properties Str-T\ornm{}-S\ornm{} and Str-D\ornm{}-I\ornm{} (analogous to Prop.~\ref{prop:str-admissibility-rr}); (ii) allow for an additive decomposition of the total variation log-odds measure, defined as TV\ornm{}$_{x_0, x_1}(\log o) = \ex[\log O \mid x_1] - \ex[\log O \mid x_0]$, into contributions from direct, indirect, and spurious log-odds effects, and their interactions (analogous to Thm.~\ref{thm:1st-tvr-decomp}). To avoid repetition, we omit the formal results for the log-odds scale.

\end{document}